\documentclass[useAMS,usenatbib]{mn2e}
\usepackage{aas_macros}
\usepackage{graphicx}
\usepackage{amssymb}
\usepackage{amsmath}

\newcommand{\Msol}{{\rm M}_\odot}

\title[Removal and mixing of the coronal gas from satellites in galaxy groups]
{Removal and mixing of the coronal gas from satellites in galaxy groups: cooling the intragoup gas}
\author[Jes\'us Zavala, Michael L. Balogh, Niayesh Afshordi and Stephen Ro]{\parbox{18cm}{Jes\'us
    Zavala$^{1,2}$\thanks{CITA National Fellow, e-mail: jzavalaf@uwaterloo.ca}, Michael L. Balogh$^{1}$,
    Niayesh Afshordi$^{2,1}$ and Stephen Ro$^{1}$\vspace{0.3cm}}\\ 
$^{1}$Department of Physics and Astronomy, University of Waterloo, Waterloo, Ontario, N2L 3G1, Canada\\
$^{2}$Perimeter Institute for Theoretical Physics, 31 Caroline St. N., Waterloo, ON, N2L 2Y5, Canada}

\begin{document}



\maketitle

\label{firstpage}

\begin{abstract}
The existence of an extended hot gaseous corona surrounding clusters, groups and massive galaxies is well
established by observational evidence and predicted by current theories of galaxy formation. When a small galaxy
collides with a larger one, their coronae are the first to interact, producing disturbances that
remove gas from the smaller system and settle it into the corona of the larger one. For a Milky-Way-size
galaxy merging into a low-mass group, ram pressure stripping and the Kelvin-Helmholtz instability are the most
relevant of these disturbances. We argue that the turbulence generated by the latter mixes the material of both
coronae in the wake of the orbiting satellite creating a ``warm phase'' mixture with a cooling time a factor of 
several shorter than that of the ambient intragroup gas. 
We reach this conclusion using analytic estimates, as well as adiabatic
and dissipative high resolution numerical simulations of a spherical corona subject to the ablation process of 
a constant velocity wind with uniform density and temperature.
Although this is a preliminary analysis, our results are promising and we speculate that the mixture could potentially trigger 
in situ star formation and/or be accreted into the central galaxy as a cold gas flow resulting in a new mode of
star formation in galaxy groups and clusters.

\end{abstract}


\section{Introduction}

The classical theory of galaxy formation establishes that gas follows dark matter in its non-linear evolution as it forms virialized
dark matter haloes. In this process, the gas is heated by shocks and adiabatic compression, acquiring the virial temperature
of the halo it has fallen into \citep{White_Rees_1978}. 
Afterwards, the gas can lose energy effectively through radiative cooling if
the cooling time is much smaller than the
free-fall time ($t_{\rm cool}<t_{\rm ff}$), producing a contraction of the gas towards the centre: a cooling flow that
fuels star formation. The critical
mass for effective cooling depends on the density, temperature and metallicity of the gas. Assuming constant
gas density and temperature, one can show that the gas in haloes with primordial gas masses $\gtrsim10^{11}$M$_{\rm \odot}$ cannot 
cool effectively \citep[e.g. see Fig. 8.6 of][]{Simon_book_2010}; this is why galaxy groups and clusters are expected to have a corona of hot
gas at the present time. 

X-ray observations of relaxed clusters and groups show indeed the presence of an extended gas corona within the 
intra-cluster and intra-group medium (hereafter called ICM in both cases), which is approximately in hydrostatic equilibrium within 
the gravitational 
potential of the cluster/group. For the most massive systems, the density profile of this diffuse
gas approximately traces that of the predicted dark matter distribution \citep[e.g.][]{Vik_2006}. Since  the dominant cooling channel in these
systems is bremsstrahlung radiation, then $t_{\rm cool}\propto T^{1/2}\rho^{-1}$ (for primordial composition) which implies that the cooling 
time scales are typically shorter in the central regions, and hence, in the absence of a balancing heat source, a cooling flow would develop 
in the centre. Close to the virial radius of the system however, the typical cooling times are larger than the Hubble time.

A hot extended corona is also predicted to exist within the haloes of massive galaxies, including those at the scale of the 
Milky-Way \citep[MW;][]{White_Frenk_1991}. In the absence of additional energy sources, and under idealized circumstances, the majority of this gas
would be expected to cool and collapse into the halo centre over a Hubble time; in contrast, roughly $20\%$ of 
the expected baryons in these haloes are observed to be in a cold phase (stars and cold gas, see e.g. \citealt{Fukugita_2004,Fukugita_2006},
and references therein). This overcooling problem can be avoided by invoking feedback processes that prevent the gas from cooling 
catastrophically. Recent cosmological simulations that include radiative cooling, star formation and feedback predict that MW-size haloes
should have a gas corona surviving today in quasi-hydrostatic equilibrium \citep[e.g.][]{Crain_2010,voort_2011}.
It has also been argued that even without feedback, the combined effect of buoyancy and thermal conduction suppresses the thermal 
instability, and thus the formation of cold clouds, in stratified coronae in MW-size galaxies \citep{Binney_2009, Nipoti_2010}.

The presence of extended coronae has been confirmed for elliptical galaxies through X-ray observations 
\citep[see review by][]{Mathews_2003} but the evidence for MW-like galaxies is not
conclusive. Recent observations have detected diffuse X-ray emission around disc-dominated galaxies, but is approximately two orders 
of magnitude lower than the expectations based on the \citet{White_Frenk_1991} model \citep[e.g.][]{Li_2007}; supernovae heating
is advocated as the likely cause of this discrepancy \citep[e.g.][]{Crain_2010,Anderson_2011,Dai_2011}.
We note that although there is evidence for the presence of gaseous coronae around galaxies, their cosmological origin as a shock-heated
remnant is not settled. There are other possible origins such as enriched gas produced by 
galactic stellar feedback \citep[e.g.][]{Ciotti_1997,Mathews_1998}.

It is thus relevant to consider in detail the impact of the gaseous coronae on the process of galaxy
formation, particularly during galaxy mergers.
In these events, the gas in the merging galaxies is subject to different processes that remove  
gas from the satellite: i) ram pressure stripping removes the most loosely bound material \citep{GG_1972} creating a
wake that trails behind the satellite (recent observations show evidence of such wakes in galaxies falling into clusters, e.g.
\citealt{Sun_2010}, but the stripped gas in the observed cases is the cold ISM of the galaxy, not 
its corona); ii) thermal conduction causes evaporation of the colder gas in the satellite \citep{Cowie_1977}; 
iii) if the flow around the boundary interface is laminar then viscosity will cause a drag that will remove gas \citep{Nulsen_1982}; 
iv) fluid instabilities such as the Kelvin-Helmholtz (KH) instability caused by the velocity shear between the layer interface also
lead to gas removal. 

All these processes have been studied in the past and for the case of galaxies merging into
clusters, particular attention has been given to ram pressure stripping using analytic and numerical methods 
\citep{Mori_2000,Heinz_2003,McCarthy_2008}. Most of these studies focus on the impact of the removal of gas in the
star formation efficiency of the satellite: by completely stripping the gas corona, a fuel reservoir
is removed, resulting in a quenching of star formation relative to a case where the galaxy evolves in
isolation (this process is usually called ``strangulation''). Recent analyses show that a substantial fraction of the
corona can actually survive if the satellite is massive enough, and through subsequent radiative cooling, can contribute to 
star formation \citep{McCarthy_2008,Moster_2011}. 

The fate of the stripped gas has not been analysed in detail however. 
As mentioned before, the gas from the satellite removed by ram pressure will form an elongated wake trailing the satellite.
Due to the gravity of the satellite, a similar wake will be formed with gas from the ICM that gets gravitationally focused 
\citep{Stevens_1999,Sake_2000}. 
Thus, in general we would expect a wake composed of material from both the ICM and the corona of the satellite, implying that both gaseous
components mix during the removal processes. This mixing is particularly efficient during the turbulence caused by 
KH instabilities. 
If the gas removal rate is large and the mixing process
is efficient then the mixture will have a temperature that might be significantly lower than the temperature of the ICM. This
would imply that the cooling time in the wake is lower than that in the ambient medium, making the mixture a potential
fuel source for star formation. This idea has been proposed recently by \citet{Marinacci_2010} in the context of a mode of gas accretion
triggered by the interaction of H$_{\rm I}$ clouds expelled from galactic discs in a galactic fountain, and the corona of the
galaxy \citep[see also][]{HP_2009}. In this paper we argue that the same mechanism is at play when the coronae of the merging
galaxies interact.

Most of the hydrodynamical simulations of galaxy formation carried out to date have used the smoothed particle 
hydrodynamics (SPH) technique \citep[for a review see][]{Springel_2010}. This is a formulation that has excellent conservation properties
in terms of mass, energy, momentum and entropy and it is fully Galilean invariant. 
However, SPH considerably suppresses the formation of turbulence triggered
by instabilities and thus it is not able to account for gas mixing \citep{Agertz_2007,Bauer_2011}.
Adaptive Mesh Refinement (AMR) codes based on an Eulerian formulation are the other method currently used for
simulations of galaxy formation. Although they suffer their own shortcomings (like the lack of Galilean invariance), perhaps
the most important difference with SPH methods is precisely that of the treatment of turbulence and mixing \citep{Mitchell_2009},
which are not suppressed in AMR. 

In this paper we take a first step in investigating the effect of mixing in the interaction of coronae in galaxy mergers, focusing
on the formation of the trailing wake. For 
this purpose we start by revisiting analytically the most important gas removal processes 
and estimate the properties of the resultant gas mixture (Section \ref{theo}). We then verify these estimates using 3D adiabatic
(non-radiative), as well as dissipative simulations of the motion of a spherical corona through a constant density ICM using the 
AMR code FLASH (Section \ref{sims}). Our results are presented in Section \ref{results} and finally a discussion and our
conclusions are given in Sections \ref{sec_discussion} and \ref{conclusions}. Although our simulations lack several important physical 
processes that could impact the results, our objective here is to make an initial assessment of whether or not the removed coronal gas 
could continue to be a source of star formation.

\section{Theoretical expectations}\label{theo}

\begin{table*}
 \begin{minipage}{100mm}
   \caption{Typical values for the global properties of a MW-size galaxy, a low-mass group and a cluster.}
   \bigskip \label{Table_1}
   \begin{tabular*}{\textwidth}{l*{6}{c}}
  \hline
   System & $M_{\rm v}(\Msol)$ &
 $r_{\rm v}({\rm kpc})$ & $V_{\rm v}({\rm km/s})$ & $T_{\rm v}({\rm K})$ & $c_{\rm s}({\rm km/s})$\footnote{The local sound speed 
   is given by $c_{\rm s}^2=\gamma P/\rho$ with $\gamma=5/3$.} \\
 \hline
 MW-size~galaxy  & $10^{12}$  & 200 & 146 & $7.7\times10^5$ & 135 \\
 low-mass~group  & $10^{13}$ & 430 & 315 & $3.6\times10^6$ & 291 \\
 cluster  & $10^{14}$ & 931 & 680 & $1.7\times10^7$ & 627 \\
\hline
\end{tabular*}
\end{minipage}
\end{table*}

In this section, we present analytic estimates of what might happen to the gas in the corona of
a satellite as it merges with that of a group/cluster. The corona is embedded in a spherical 
dark matter halo which is assumed to have a radial density distribution given by the Navarro-Frenk-White (NFW) 
profile \citep{NFW_1996,NFW_1997}\footnote{The most 
recent N-body simulations of structure formation suggest that the density profile is actually shallower than
NFW towards the centre with an Einasto profile giving a better fit \citep{Springel_2008}. However, since 
we are not interested in the central regions, the NFW profile is an appropriate choice.}, which in terms of the
dimensionless variable $s=r/r_{\rm v}$, where $r_{\rm v}$ is the virial radius, can be written as:
\begin{equation}\label{nfw}
  \rho(s)=\frac{\rho_{\rm crit}\delta_{\rm c}}{cs(1+cs)^2},
\end{equation}
where $c$ is the concentration parameter, $\rho_{\rm crit}$ is the critical density and $\delta_{\rm c}$ is a function of the concentration:
\begin{equation}
  \delta_{\rm c}=\frac{\Delta}{3}\frac{c^3}{{\rm log}(1+c)-c/(1+c)}.
\end{equation}
The virial radius $r_{\rm v}$ is defined as the radius where the mean density of the halo is equal to $\Delta$ times 
the critical density: $M_{\rm v}/(4\pi r_{\rm v}^3/3)=\Delta\rho_{\rm crit}$, where $M_{\rm v}$ is the total mass within $r_{\rm v}$. The 
value of the density contrast $\Delta$ is conventional and varies in the literature, but we use $\Delta=200$ for this work. 
The concentration correlates strongly with virial mass and redshift \citep[e.g.][]{Gao_2008}. 
When a numerical value is needed, we adopt the mass-concentration relation found in \citet{Neto_2007} based on the 
Millennium simulation of structure formation \citep{Springel_2005}.

We further assume that the dark matter haloes contain a distribution of gas that also follows the NFW profile but
normalised by the universal fraction of baryons to dark matter $f_{\rm b}=\Omega_{\rm b}/(\Omega_{\rm m}-\Omega_{\rm b})$. Whenever it is needed, 
we take $\Omega_{\rm m}=0.25$ and $\Omega_{\rm b}=0.045$ as the values for the contributions from matter and baryons to the total 
energy density of the Universe, respectively.

The pressure profile is obtained by assuming that the coronal gas is in hydrostatic equilibrium: 
\begin{equation}\label{hse}
  \frac{dP}{ds}=-\rho_{\rm gas}\frac{d\phi}{ds},
\end{equation}
where $\phi(s)$ is the total gravitational potential (dark matter + gas). To obtain a boundary condition for 
Eq.~(\ref{hse}), we note that the virial velocity of the dark matter halo $V_{\rm v}^2=GM_{\rm v}/r_{\rm v}$ determines a characteristic
temperature for the corona $T_{\rm v}=\mu m_{\rm H}V_{\rm v}^2/(2k_{\rm B})$, where $\mu$ is the mean molecular weight of the
gas ($\mu^{-1}=1.71$ for gas with primordial composition), $m_{\rm H}$ is the mass of a hydrogen atom and $k_{\rm B}$ is the 
Boltzmann constant. This characteristic value, and also the assumption of hydrostatic equilibrium, are motivated by
the formation of the corona when the gas fell into the halo and got shock-heated 
to temperatures of the order of $T_{\rm v}$. The boundary condition for Eq.~(\ref{hse}) is then set
by $P_{\rm v}$ given by the value of the virial temperature of the halo and the equation of state (EoS), assumed
to be that of an ideal gas. Since $T_{\rm v}$ is actually close to the
peak of the temperature profile, we apply a small correction (a factor of $\sim3/4$) to the
value of $P_{\rm v}$ to obtain a temperature profile that decays as a power law at large radii, which is a behaviour
seen in observations of nearby galaxy clusters and groups \citep[e.g.][]{Vik_2005}, as well
as in full hydrodynamical simulations of such systems \citep[e.g.][]{Borgani_2004}.

Unless otherwise stated, we assume that the satellite halo enters the ICM 
with a relative velocity $v_{\rm sat}$ which is of the order of the virial velocity of the host $V_{\rm v,host}$. 
Throughout this section, we compare the relevance of different
processes during the interaction of the hot coronae in two cases: a MW-size galaxy falling into a low-mass group (galaxy-group
collision) and the same galaxy falling into a cluster (galaxy-cluster collision). The merger mass ratio in the cases
of interest is $\leq10$, so the gravitational potential of the most massive system is not significantly 
affected. The characteristic parameters of these three systems appear in Table \ref{Table_1}.

\subsection{Removal of gas from the satellite}

\subsubsection{Ram Pressure Stripping (RPS)}\label{RPS_sec}

As the satellite moves through the ICM, the ram pressure ($P_{\rm ram}\equiv\rho_{\rm host}v_{\rm sat}^2$) imparted to its corona 
will be sufficient to remove gas from the satellite if it can overcome its gravitational restoring force per 
unit area \citep[][based on \citealt{GG_1972}]{McCarthy_2008}:
\begin{equation}\label{RPS}
  P_{\rm ram}>\alpha_{\rm g}\frac{Gm_{\rm tot}(<r)\rho_{\rm sat}(r)}{r},
\end{equation}
where $\rho_{\rm sat}(r)$ is the gas density profile of the corona of the satellite, $m_{\rm tot}(<r)$ its total mass profile 
(dark matter + gas), and $\alpha_{\rm g}$ is a geometric factor of $\mathcal{O}(1)$ depending on 
the total mass and gas density profiles of the satellite; for instance, $\alpha_{\rm g}=\pi/2$ for a singular isothermal sphere. The 
radius where the lhs and rhs of Eq.~(\ref{RPS}) are equal is the predicted radius $r_{\rm rps}$ that the gas sphere 
will have after stripping; $m_{\rm sat}(<r_{\rm rps})$ is then the surviving mass. 

As long as the relative velocity between both systems is supersonic, the collision drives a shock that propagates 
through the satellite; otherwise a compression wave is formed that propagates with the local sound speed.
It is possible to estimate the speed 
of this shock and its subsequent propagation as a function of the radius of the satellite $v_{\rm sh}(r)$ by solving a one-dimensional 
shock tube problem, neglecting the effects of gravity (see Appendix \ref{app_1}). The characteristic 
RPS time scale is then roughly given by the time it takes for this shock to cross the diameter of the satellite:
\begin{equation}\label{t_rps}
  t_{\rm rps}\sim2\int_0^{r_{\rm v}}\frac{dr}{v_{\rm sh}(r)}.
\end{equation} 
The average mass removal rate from RPS is then given by:
\begin{equation}
  \dot{m}_{\rm rps}\sim\frac{m_{\rm sat}(<r_{\rm rps})-m_{\rm sat}(<r_{\rm v,sat})}{t_{\rm rps}}.
\end{equation}

\subsubsection{Tidal Stripping (TS)}

Tides remove gas and dark matter from the satellite as it falls into the host. 
The tidal radius $r_{\rm t}$, defined as the radius where the external 
differential tidal force from the host exceeds the binding force of the satellite, 
is approximately given by \citep[e.g.][]{Taylor_2001}:
\begin{equation}\label{tidal}
  r_{\rm t}^3=\frac{Gm_{\rm tot}(<r_{\rm t})}{w^2+G\left[2M(<r)/r^3-4\pi\rho_{\rm host}(r)\right]},
\end{equation} 
where $w$ is the angular speed of the satellite located at a distance $r$ from the host, which has a mass and density profile
$M(<r)$ and $\rho_{\rm host}(r)$ respectively. The material outside $r_{\rm t}$ is tidally removed from the system
within a time scale $t_{\rm ts}=t_{\rm orb}/A$ where $t_{\rm orb}=2\pi/w$ is the 
instantaneous orbital time and $A$ is an efficiency parameter. It is important to note that Eq.~(\ref{tidal}) is 
only a rough approximation\footnote{For instance, the tidal radius is only well-defined for circular orbits, whereas realistic
orbits are in general eccentric.}. The uncertainties associated with this equation are parameterised by the factor
$A$ that has values between 1 and 6 according to numerical simulations of dark matter
subhaloes orbiting host haloes \citep{Taylor_2001,Zentner_2005,Diemand_2007}.
We take a fiducial value of $A=3.5$ as found by \citet{Gan_2010},
and define the mass removal rate due to TS as: 
\begin{equation}\label{mdot_ts}
  \dot{m}_{\rm ts}=\frac{m_{\rm sat}(<r_{\rm t})-m_{\rm sat}(<r_{\rm v,sat})}{t_{\rm ts}}.
\end{equation}

Besides RPS and TS, the corona of the satellite can be subjected to other mass loss mechanisms which we mention
subsequently (closely following \citealt{Nulsen_1982}).

\subsubsection{Laminar viscous stripping}\label{viscous}

If viscosity is relevant in the ICM, then there will be a drag in the corona of the satellite resulting in a gas mass loss 
which is roughly given by (ignoring gravity):
\begin{equation}\label{visc}
  \dot{m}_{\rm visc}\approx\left(\frac{12}{\rm Re}\right)\pi r_{\rm sat}^2\rho_{\rm host}v_{\rm sat},
\end{equation}
where $r_{\rm sat}$ is the instantaneous maximum radial extent of the corona of the satellite and Re is the Reynolds number:
\begin{equation} 
{\rm Re}=2.8\left(\frac{r_{\rm sat}}{\lambda_{\rm host}}\right)\left(\frac{v_{\rm sat}}{c_{\rm s, host}}\right),
\end{equation}
with $\lambda_{\rm host}$ being the effective 
mean free path of ions in the ICM. In the absence of magnetic fields and for Coulomb collisions this 
is given by \citep{Spitzer_1956}:
\begin{equation}
  \lambda_{\rm host}\approx11~{\rm kpc}\left(\frac{T_{\rm host}}{10^8 {\rm K}}\right)^2
  \left(\frac{\rho_{\rm host}/m_{H}}{10^{-3}{\rm cm^{-3}}}\right)^{-1}.
\end{equation}

For Eq.~(\ref{visc}) to be applicable, the gas flow needs to be laminar around the boundary of the satellite, i.e., ${\rm Re}\leq30$, and
also $\lambda_{\rm host}\leq r_{\rm sat}$. Since $T^2_{\rm host}/\rho_{\rm host}\propto r^{1.9}$ for the NFW profile (and for the
temperature given by the EoS assuming hydrostatic equilibrium), $\lambda_{\rm host}(r)\leq \lambda_{\rm host}(r_{\rm v,sat})$. At the
virial radius of the host, the mean free path is of $\mathcal O(1~{\rm kpc})$ and since $v_{\rm sat}/c_{\rm s,host}\gtrsim1$ (the 
initial shock is transonic), the Reynolds number is then very large and viscosity is irrelevant (see Table \ref{Table_2}). Due 
to RPS and TS, $r_{\rm sat}$ will be reduced as the satellite spirals inwards to the centre of the host but the ratio 
$r_{\rm sat}/\lambda_{\rm host}$ will certainly remain very large. The changes in the ratio $v_{\rm sat}/c_{\rm s,host}$ will depend
strongly on the orbit but will likely remain larger than 1, increasing towards pericentre, and dropping below 1 near apocentre. 
If apocentre is near the virial radius of the group (cluster) and the radius of the satellite has been reduced by a 
significant factor, then Re could drop below 30 in the cluster environment. This is unlikely to happen in the group environment 
because the mean free path of the ions in the ICM is an order of magnitude smaller than in the cluster environment. We 
neglect viscous stripping in the remainder of this work.

\begin{table*}
 \begin{minipage}{150mm}
   \centering
   \caption{Typical values for the parameters defining different processes of gas removal for a MW-size galaxy falling 
   into a group/cluster (see Table \ref{Table_1}). Times are given in Gyr and mass loss rates in ${\rm \Msol/yr}$.}
   \bigskip \label{Table_2}
  \begin{tabular*}{\textwidth}{l*{11}{c}}
  \hline
   Environment & $r_{\rm rps}/r_{\rm v}$ & $t_{\rm rps}$ & $\dot{m}_{\rm rps}$ & $r_{\rm t}/r_{\rm v}$ & 
   $t_{\rm ts}$ & $\dot{m}_{\rm ts}$ & $\lambda_{\rm host}/r_{\rm v}$ & ${\rm Re}$ & $t_{\rm KH}$ &
   $\dot{m}_{\rm KH}$ & $\dot{m}_{\rm evap}$\\
 \hline
 galaxy-group  & 0.78  & 2.1 & 15 & 0.65 & 2.4 & 22 & $7\times10^{-4}$ & 4800 & 1.2 & 39 & 0.5 \\
 galaxy-cluster & 0.45  & 1.9 & 50 & 0.65 & 2.4 & 22 & $1.4\times10^{-2}$ & 250 & 0.56 & 53 & 23 \\
\hline
\end{tabular*}
\end{minipage}
\end{table*}

\subsubsection{Evaporation due to thermal conduction}

The rate of evaporated mass due to thermal conduction from a spherical cloud embedded in a hot tenuous gas, neglecting 
gravity, radiation, ionization and magnetic fields, is approximately given by \citep{Cowie_1977}:
\begin{equation}\label{evap}
  \dot{m}_{\rm evap}\approx4\pi r_{\rm sat}^2\rho_{\rm host}c_{\rm s,host}\phi_{\rm s}F(\sigma_0),
\end{equation}
where $\phi_{\rm s}$ is a parameter of $\mathcal O(1)$ and $\sigma_0\equiv1.84\lambda_{\rm host}/r_{\rm sat}\phi_{\rm s}$ determines when the 
regime of classical conductivity still applies. This happens when the mean free path is short compared 
to the temperature scale height $L_{\rm T}=T/\vert \nabla T\vert$ between the boundary of the cloud and the hot gas surrounding it 
($\lambda_{\rm host}\ll L_{\rm T}$); in this regime, $F(\sigma_0)=2\sigma_0$. 
For the set of interacting systems we are considering, $r_{\rm sat}/\lambda_{\rm host}>\mathcal O(100)$, thus, 
$\sigma_0<\mathcal O(0.01)$, which makes the mass loss rate due to thermal conduction much lower than that of the other
processes for the group environment (see Table \ref{Table_2}). For the cluster environment, evaporation is still subdominant
but it cannot be entirely neglected.

\subsubsection{Turbulent Stripping (Kelvin-Helmholtz Instability)}

If the inertial forces in the surface layers between the two gases are larger than those related to viscosity (i.e. if 
${\rm Re}>30$), then the velocity shear between the layers will drive KH instabilities. The collision
between the coronae of the satellite and the ICM also drives a reverse shock that propagates into the ICM (contrary to
the forward shock propagating into the satellite and related to RPS). Since, typically, $v_{\rm sat}\sim c_{\rm s,host}$ at infall, 
this shock is generally weak. The post-shock flow in the vicinity of the boundary of the satellite, where the instabilities 
develop, will then be subsonic (propagating at a speed $v_{\rm ps}$, see Eq.~A2). Under this assumption, and taking into 
account the compressibility of the gas flow, 
the mass loss rate for complete stripping due to KH instabilities is:
\begin{equation}\label{mdot_khi}
  \dot{m}_{\rm KH}\sim\pi r_{\rm sat}^2\rho_{\rm host}v_{\rm ps}.
\end{equation}
In this equation, the dominant wavelength of the perturbations 
is of the order of the radius of the corona of the satellite. This is because the stripping process begins with perturbations that
have shorter wavelengths; as they develop, they smooth the velocity gradient between the layers making
the interface thicker. This damps modes with wavelengths smaller than the thickness of the interface,
while larger wavelength modes continue growing, stripping more material and making the interface thicker. 
This process continues until the layer has a thickness of order $r_{\rm sat}$. 

As we mentioned before, the velocity flow near the layer is expected to be subsonic. We
estimate the values of $v_{\rm ps}$ by solving a one dimensional shock tube problem (see Appendix \ref{app_1}) using the
value of the hydrodynamical variables at the beginning of the interaction (assuming initial pressure equilibrium 
between both systems). We find $v_{\rm ps}\sim208~(260) {\rm km/s}$ for the group (cluster) environments resulting in
mass loss rates which are of the order of those by RPS and TS (see Table \ref{Table_2}). The time scale for the formation of 
the largest wavelength instabilities can be roughly estimated for an ideal case where the fluids are
incompressible, the surfaces are flat, gravity is neglected, and the overdensity near the boundary is in the linear 
regime: $\delta=\rho_{\rm sat}/\rho_{\rm host}-1\ll1$. In this case \citep[e.g., ][]{Drazin_1981}:
\begin{equation}\label{t_khi}
  t_{\rm KH}\sim\frac{2r_{\rm sat}}{v_{\rm sat}},
\end{equation}
which is of the order of the sound crossing time.

Gravity and density stratification of the gas are stabilizing effects acting against the development of KH instabilities. By assuming 
that the interface has a sharp density discontinuity between the gas in the satellite $\rho_2$ and the ICM gas $\rho_1$, 
\citet{Murray_1993} found that the dominant wavelength for gas ablation, $k=2\pi/r_{\rm sat}$, is stable if:
\begin{equation}
  g>\frac{2\pi\rho_1\rho_2v_{\rm sat}^2}{r_{\rm sat}(\rho_2^2-\rho_1^2)},
\end{equation}
where $g=Gm_{\rm tot}(<r_{\rm sat})/r_{\rm sat}^2$ is the gravitational acceleration at the interface. This condition can be
rewritten in terms of the gravitational restoring pressure (Eq.~\ref{RPS}):
\begin{equation}\label{res_khi}
  P_{\rm restore}(r_{\rm sat})>\frac{2\pi\alpha_{\rm g}\rho_1v_{\rm sat}^2}{1-\left(\rho_1/\rho_2\right)^2}.
\end{equation}
The rhs of Eq.~(\ref{res_khi}) is expected to be much larger than the restoring force in the outskirts of the satellite, and
thus, the corona will not be stabilized by its own self-gravity but the mass loss rate in Eq.~(\ref{mdot_khi})
is probably going to be reduced. Other physical processes, such as magnetic
fields \citep[e.g.][]{Malagoli_1996} and radiative cooling \citep[e.g.][]{Vietri_1997},
might stabilize the corona of the satellite. 

Given the parameters of the interacting galaxies, it seems likely that RPS, TS and KH instabilities will act on similar time scales, 
competing to remove the gas from the corona of the satellite. Besides gas removal, all these processes will also mix the stripped gas
with the gas in the ICM with important consequences as we discuss below.

\subsection{Cooling due to mixing}\label{sec_cooling}

As the satellite orbits the host, it loses energy and angular momentum due to dynamical friction \citep{Chan_1943} eventually 
merging completely with the host. The typical time scale for this to happen is $\sim4(23)$~Gyr for the satellite merging into 
the group (cluster) (following the study of \citealt{Boylan_2008}). This implies that the satellite will remain in orbit for 
a reasonable time before getting too strongly disrupted. Thus, the gas removed from the satellite 
will have enough time to mix with the ICM, 
changing its properties. In the following, we consider this gas mixing process, inspired by the study of \citet{Marinacci_2010}. 

The coronal gas is being removed at a rate given by the different stripping processes. 
If we take the mass loss rate by RPS as a reference, then $\dot{m}_{\rm rem}=\alpha_{\rm rem}\dot{m}_{\rm rps}$, with 
$\alpha_{\rm rem}\gtrsim\mathcal O(1)$ representing the mass loss for other processes. The removed gas 
mixes with the ICM (possibly quite efficiently due to the turbulent flow generated by KH instabilities) and falls behind the satellite
forming a wake. In the time it takes the satellite to move a distance $s$, the gas mixture 
occupies a volume $V_{\rm w}=A_{\rm w}s$, where $A_{\rm w}$ is the cross section of the wake. If we neglect 
the compressibility
of the gas, then the mass from the corona and the ICM contained within $V_{\rm w}$ is: 
$M_{\rm w}=M_{\rm host, w}+M_{\rm sat, w}=A_{\rm w}s\rho_{\rm host}+\dot{m}_{\rm rem}s/v_{\rm sat}$. If the gases in the corona and the ICM 
have the same composition, then the temperature of the mixture is just:
\begin{eqnarray}\label{eq_Tm}
  T_{\rm m}&=&\frac{M_{\rm host,w}T_{\rm host}+M_{\rm sat,w}T_{\rm sat}}{M_{\rm host,w}+M_{\rm sat,w}}=\nonumber\\
 &&\frac{A_{\rm w}\rho_{\rm host}T_{\rm host}+\dot{m}_{\rm rem}T_{\rm sat}/v_{\rm sat}}{A_{\rm w}\rho_{\rm host}+\dot{m}_{\rm rem}/v_{\rm sat}}.
\end{eqnarray}
Note that thermal conduction is necessary to 
reach this temperature, but this can happen efficiently on small scales due to turbulent mixing.
Depending on the relative difference between the gas from the satellite being deposited in the wake $\dot{m}_{\rm rem}/v_{\rm sat}$
to that of the ICM $A_{\rm w}\rho_{\rm host}$, the temperature of the mixture can decrease significantly relative to $T_{\rm host}$, 
down to a minimum of $T_{\rm sat}$. 

The characteristic scale for the cross section of the wake is given by a combination of the gravity of the satellite, 
the inertia of the stripped gas and turbulent effects. For simplicity, we parametrize it as $A_{\rm w}=\beta\pi r_{\rm s}^2$, where 
$r_{\rm s}$ is the scale radius of the halo of the satellite: $r_{\rm s}=r_{\rm v, sat}/c$ (with $c$ being the concentration of the halo),
and $\beta$ is a free parameter 
which in the absence of turbulence is mainly set by the velocity of the satellite\footnote{This cross section seems 
to be a good guess according to the results we found for a low resolution simulation analogous to the non-radiative simulation
described in Section \ref{sims} for a subsonic collision ($v_{\rm sat}=150$km/s). We found that the cross-sectional radius of 
the wake separating the gas in the ICM that flows without mixing with the one that mixes is $\sim r_{\rm s}$, i.e., $\beta=1$. For higher velocities, 
$\beta$ is likely to increase asymptotically to $(r_{\rm v}/r_{\rm s})^2$.}. We note that in reality, the gas mixes across layers of different sizes
in the turbulent flow creating a broad distribution of $A_{\rm w}$ values. 

The time scale for gas to lose its thermal energy due to radiative losses is given by:
\begin{equation}
  t_{\rm cool}=\frac{3/2nk_BT}{n_{\rm e}n_{\rm H}\Lambda_{\rm cool}(T,Z)},
\end{equation}
where $n=\rho_{\rm host}/(\mu m_{\rm H})$, $n_{\rm e}$ and $n_{\rm H}$ are the number densities for
electrons ($n_{\rm e}=0.52n$) and hydrogen, respectively; and $\Lambda_{\rm cool}(T,Z)$ is the cooling function that depends on temperature
and metallicity. We obtain the cooling function from the tables given by \citet{Gnat_2007}.

\begin{figure}
\center{
\includegraphics[height=8.0cm,width=8.0cm]{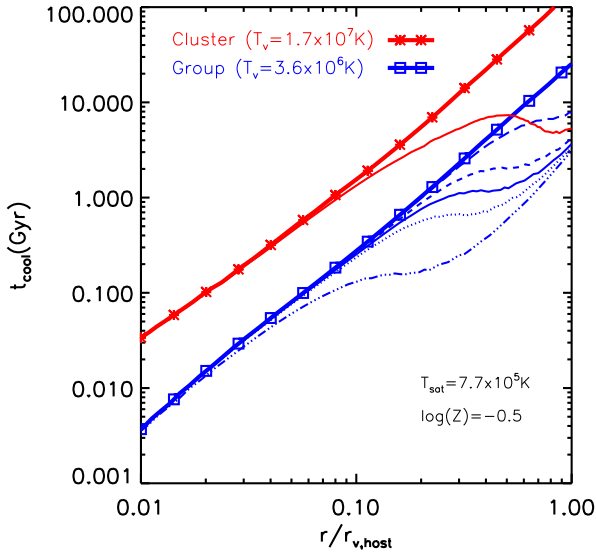}
}
\caption{Cooling time as a function of radius for a group/cluster with an NFW density profile under hydrostatic equilibrium. The
thick solid lines are for an isolated system; the thin solid lines show what would happen if gas in the ICM at different radii 
mixed with coronal gas removed from an infalling satellite ($T_{\rm sat}=7.7\times10^5$K) moving with a constant speed equal to the 
virial velocity of the group/cluster (the cooling time refers in this case to the gas in the mixture). The gas is removed at a 
constant rate $\dot{m}_{\rm rem}=\alpha_{\rm rem}\dot{m}_{\rm rps}$ 
and mixes with the ICM in a region with a cross sectional area $A_{\rm w}=\beta\pi r_{\rm s}^2$. The thin solid lines are for 
$\alpha_{\rm rem}=\beta=1$. The dashed (dotted) lines are for $\alpha_{\rm rem}=1/2(2)$ and the long-dashed (dotted-dashed) lines are
for $\beta=10(1/10)$.} 
\label{cool_mix} 
\end{figure}

Fig.~(\ref{cool_mix}) shows the cooling time for the gas mixture as a function of the distance to the centre of the group (cluster) 
assuming that the ICM in the host mixes with gas from the satellite with a single temperature ($T_{\rm sat}=T_{\rm v,sat}$).
We also assume that the density of the mixture is the same as the density of the host\footnote{Note that it is perhaps more
reasonable to assume that the mixture will have the same pressure as the ambient ICM; in this case, its density will be higher
and thus its cooling time will be lower than the one we have estimated.}, and that the 
mass loss rate and velocity of the satellite remain constant with radii. 
We have used a metallicity of log$_{10}(Z/Z_{\odot})=-0.5$, which is representative of the values found in 
groups and clusters \citep{Rasmussen_2009}. The thick solid lines show the reference cases where there is no mixing, in blue
and red for the group and cluster, respectively. The thin blue and red lines show the case when mixing is effective, with 
$\alpha_{\rm rem}=\beta=1$ for the group and cluster, respectively. Only for the group, we show lines where the ratio 
$\dot{m}_{\rm rem}/v_{\rm sat}$ is decreased (increased) by a factor of 2,  dashed (dotted) blue, and for changes in the parameter 
$\beta$ increasing (decreasing) it by a factor of 10, long-dashed (dotted-dashed) blue. The former of these explores variations
in the efficiency of gas removal, and the latter on the effective size of the region where material mixes.

The regions where the different assumptions made to arrive at Fig.~(\ref{cool_mix}) are likely to be true, are those close to the
outskirts of the host. We see that for $r/r_{\rm v, host}\sim0.6$ the cooling time of the mixed material
is of  $\mathcal O(1$~Gyr) for the group environment, 
which is an order of magnitude lower than the cooling time of the ambient ICM. For the cluster environment, the reduction in the
cooling time is also quite substantial but certainly above a few Gyrs. 

If a volume element of gas can cool down quickly enough by radiating its thermal energy, then, in the absence of any source
of heating, it will lose all its pressure support and collapse in a free-fall time:
\begin{equation}
  t_{\rm ff}=\sqrt{\frac{3\pi}{32G\rho_{\rm local}}}.
\end{equation}
The condition $t_{\rm cool}=t_{\rm ff}$ separates the gas clouds that can cool down effectively 
from those that cannot. The local free-fall time for gas at $r\sim0.6r_{\rm v, host}$ in the group(cluster) environment is $\sim4$Gyr. 

Therefore, from the theoretical expectations described in this section we conclude that: {\it the coronal gas removed from the infalling 
satellite will mix with the ICM of the host, creating a mixture with a cooling time potentially lower than the local free-fall time. The net 
effect is an increase in the cooling efficiency of the gas mixture trailing behind the merging satellite, relative to the ambient ICM}. 
Of course, the gas mixture will have a cooling time which is larger than that of the initial corona.

Note that we have chosen a mass for the MW-size galaxy which is at the low end of the range of Milky-Way halo masses allowed by 
current estimates \citep[see e.g. section 5.1 of][]{Boylan_2012}, which results in a virial temperature for the corona that is relatively low
compared to other estimates (e.g. \citealt{Fukugita_2006} estimate $T_{\rm v}=1.4\times10^6$K since they assume $M_{\rm v}=2.6\times10^{12}$M$_{\odot}$).
A hotter corona would result in a reduction of the cooling efficiency of the mixed ICM gas. We have also
chosen a low-mass group ($10^{13}$~M$_{\odot}$), corresponding to a galaxy-group merger mass ratio of $1/10$. Mergers with lower mass ratios are typically
more common in a cosmological setting \citep[see e.g.][]{Fakhouri_2010}, and as implied by Fig.~(\ref{cool_mix}) for the case of the galaxy-cluster
collision, a merger with a group of higher mass (virial temperature) would increase the temperature of the gas mixture. Nevertheless, 
the $1/10$ galaxy-group merger we are taking as a reference should be common enough to be of interest. We leave the analysis in a full 
cosmological setting for a future work and concentrate in the rest of this paper on the formation and properties
of the wakes produced by the galaxy-group interaction.

In the previous analysis, we neglected the compression of the gas during mixing, which could in principle reduce the 
cooling time and the free fall time even further. However, since an adiabatic compression would also increase the temperature 
of the gas (heating the mixture instead of cooling it), it might be that the two effects cancel each other. We
answer this question in the following section. 

It is in principle possible that a fraction of the gas trailing behind the satellite is gravitationally re-accreted. 
However, we find that for the parameters in Table \ref{Table_1}, 
the presence of the satellite in the environment of the group(cluster) results in a Bondi radius $\lesssim53(11)$kpc, 
which is smaller than the virial radius of the satellite. 
Since RPS happens fairly rapidly, the effective radius of the corona is 
quickly reduced while the mass of the dark matter satellite remains fairly the same at first since TS is expected to act more 
slowly (see Table \ref{Table_2}). Nevertheless, the stripping of the gas is not enough to reduce $r_{\rm sat}$ to the Bondi radius. 
For this to happen in the group (cluster) medium, the stripping efficiency would need to be much higher, followed by a period of 
relatively low orbital velocities (low compared to the virial velocity of the host). The satellite could then potentially 
accrete material from the ICM.
Although this situation is not common
for the systems we are studying, in the case of the galaxy-group collision, a ``fine-tuned'' orbit could potentially have a 
Bondi radius smaller than the extent of the corona of the satellite near apocentre.

\section{Simulations}\label{sims}

We use the Eulerian code FLASH \citep{Fryxell_2000}, that uses an AMR 
technique to solve the hydrodynamical equations for the fluid. With this technique, FLASH is able to increase the resolution of the 
grid solver in the regions of most interest while keeping a coarser grid elsewhere. FLASH also includes the possibility of adding 
particles to the simulation in two varieties: {\it active } and {\it passive}. The former have mass and participate in the dynamics 
of the system, and we use them to simulate the dark matter halo. 
Passive particles have no mass and do not contribute to the dynamics, they are Lagrangian tracers of the flow and are useful
to study the degree of mixing. The code uses a particle-mesh method to map the quantities from the mesh into
the particles and vice versa. The Euler equations are solved using a piecewise-parabolic method \citep{Colella_1984} and the gravitational 
potential is solved using a parallel multigrid solver \citep{Ricker_2008}.

\subsection{Initial conditions and setup}\label{IC}

The satellite consists of a spherical distribution of collisionless dark matter and gas with an
ideal equation of state. Both of these fluids have a density profile given by Eq.~(\ref{nfw}).
Dark matter is simulated with a set of $N$ particles with their positions and velocities
given by an algorithm based on the method presented in \citet{Kazan_2004}. To assign the position of each particle, we
randomly draw a value of the enclosed mass $M(<r)$ in the interval $[0,M_{\rm v}]$ and then invert the 
function $M(<r)$ to find the distance of that particle to the centre of the halo. 
The velocities of the dark matter particles are assigned from their actual energy distribution 
function given by the Eddington formula \citep[e.g.][]{Binney_1987}:
\begin{equation}\label{eddington}
  f(\varepsilon)=\frac{1}{\sqrt{8}\pi^2}\left[\int_0^\varepsilon\frac{d^2\rho}{d\Psi^2}\frac{d\Psi}{\sqrt{\varepsilon-\Psi}}+
  \frac{1}{\sqrt{\varepsilon}}\left(\frac{d\rho}{d\Psi}\right)_{\Psi=0}\right],
\end{equation}
where $\Psi=-\phi$ is the relative gravitational potential and $\varepsilon$ is the relative energy $\varepsilon=\Psi-v^2/2$. We construct
a table with values of $\varepsilon$ and $f(\varepsilon)$ that allows us to linearly interpolate for any value of the relative energy. 
The magnitude of the velocity of each particle is then drawn at random from the distribution function 
$2\left[\Psi(r)-\varepsilon\right]^{1/2}f(\varepsilon)$ using a rejection sampling method \citep{von_1951}. 
The direction of the position and velocity vectors is chosen at random from the unit sphere.
This method ensures that the dark matter halo is a system in
equilibrium (except of course from imbalances caused by discreteness effects).

\begin{figure*}
\begin{tabular}{|@{}l@{}|@{}l@{}|}
\includegraphics[height=9.2cm,width=4.0cm,angle=90,trim=4.6cm 0cm -0.9cm 0cm, clip=true]{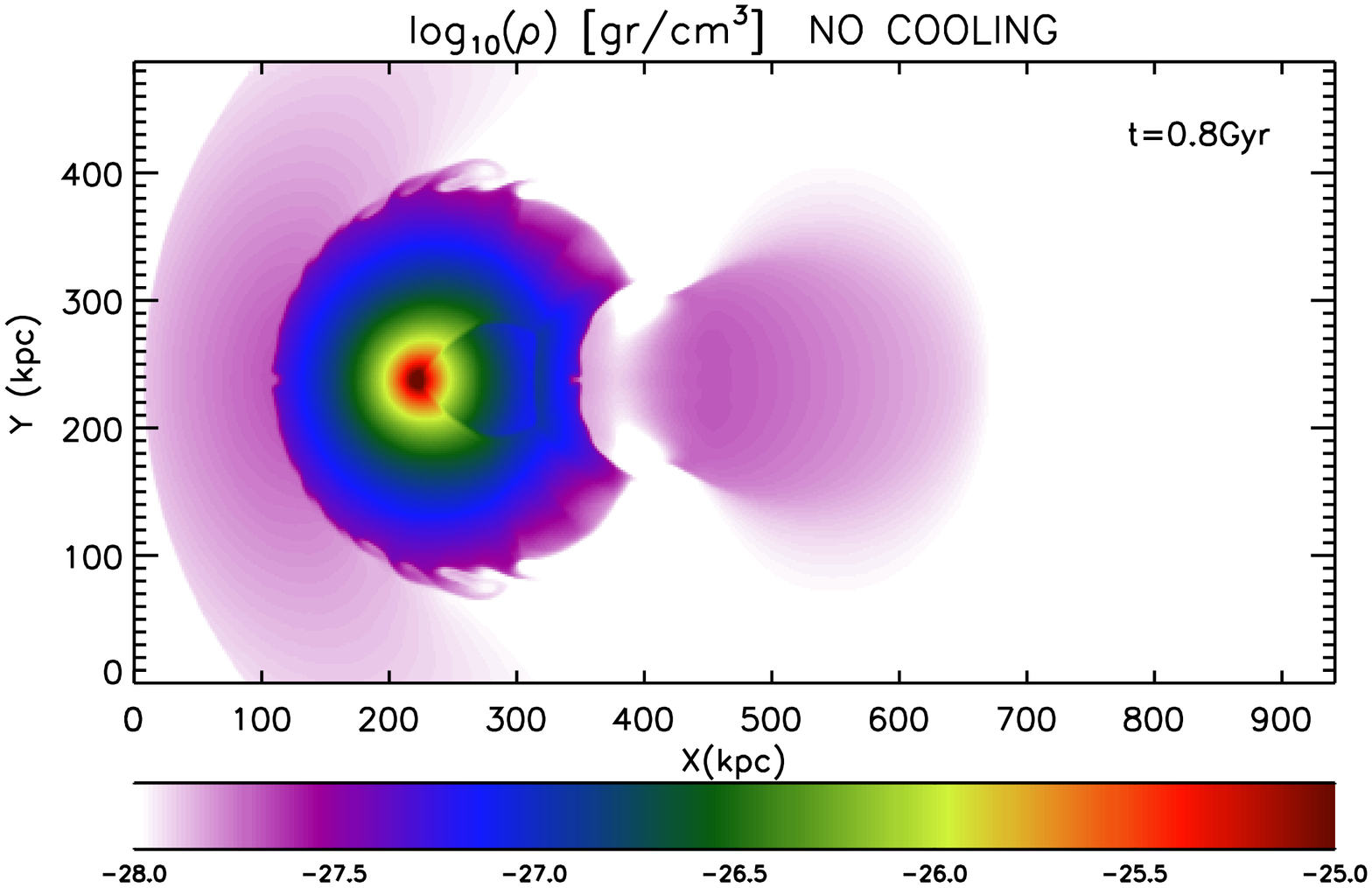} & 
\includegraphics[height=9.2cm,width=4.0cm,angle=90,trim=4.6cm 0cm -0.9cm 0cm, clip=true]{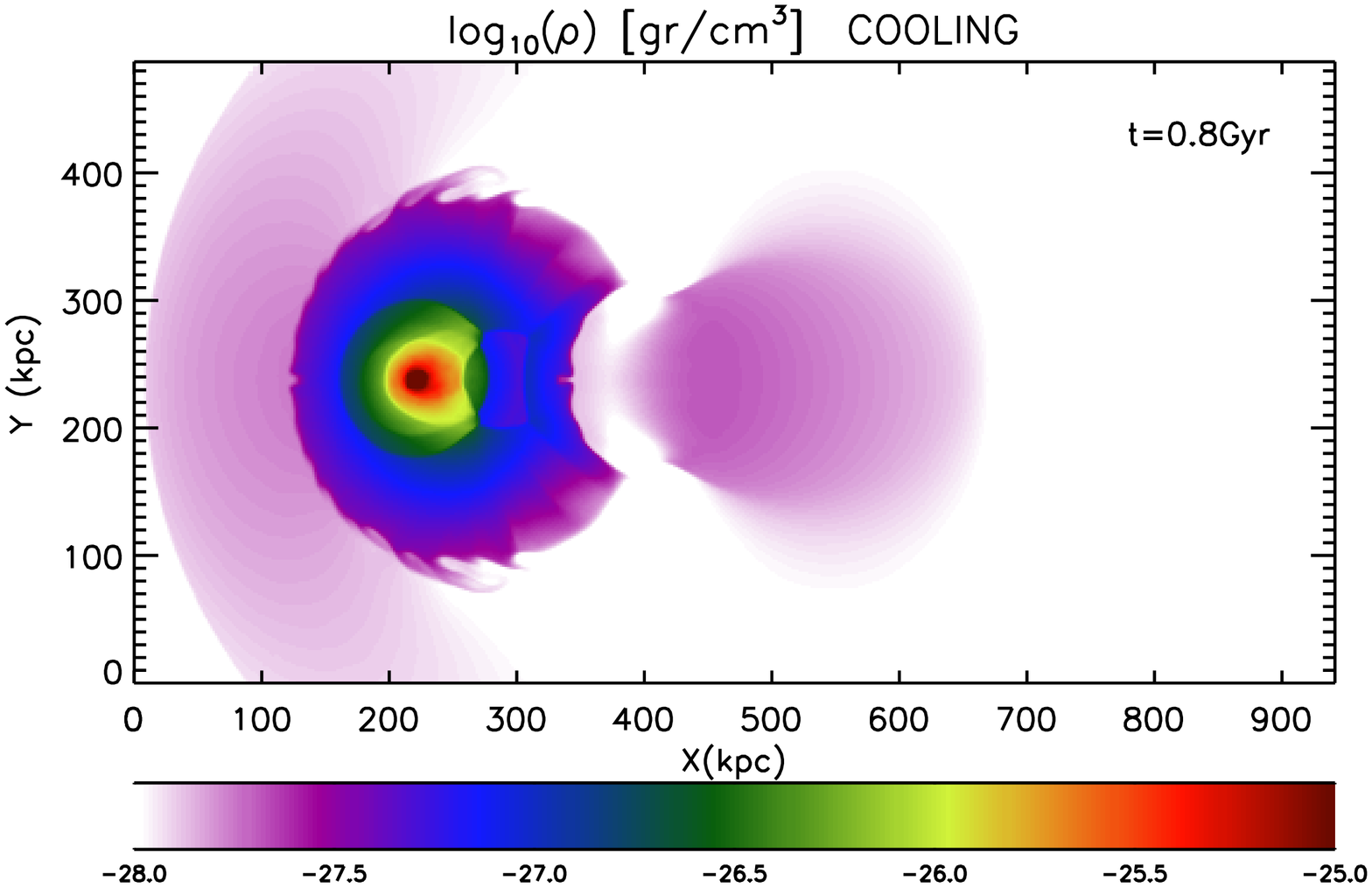} \\
\includegraphics[height=9.2cm,width=3.68cm,angle=90,trim=4.6cm 0cm 0.1cm 0cm, clip=true]{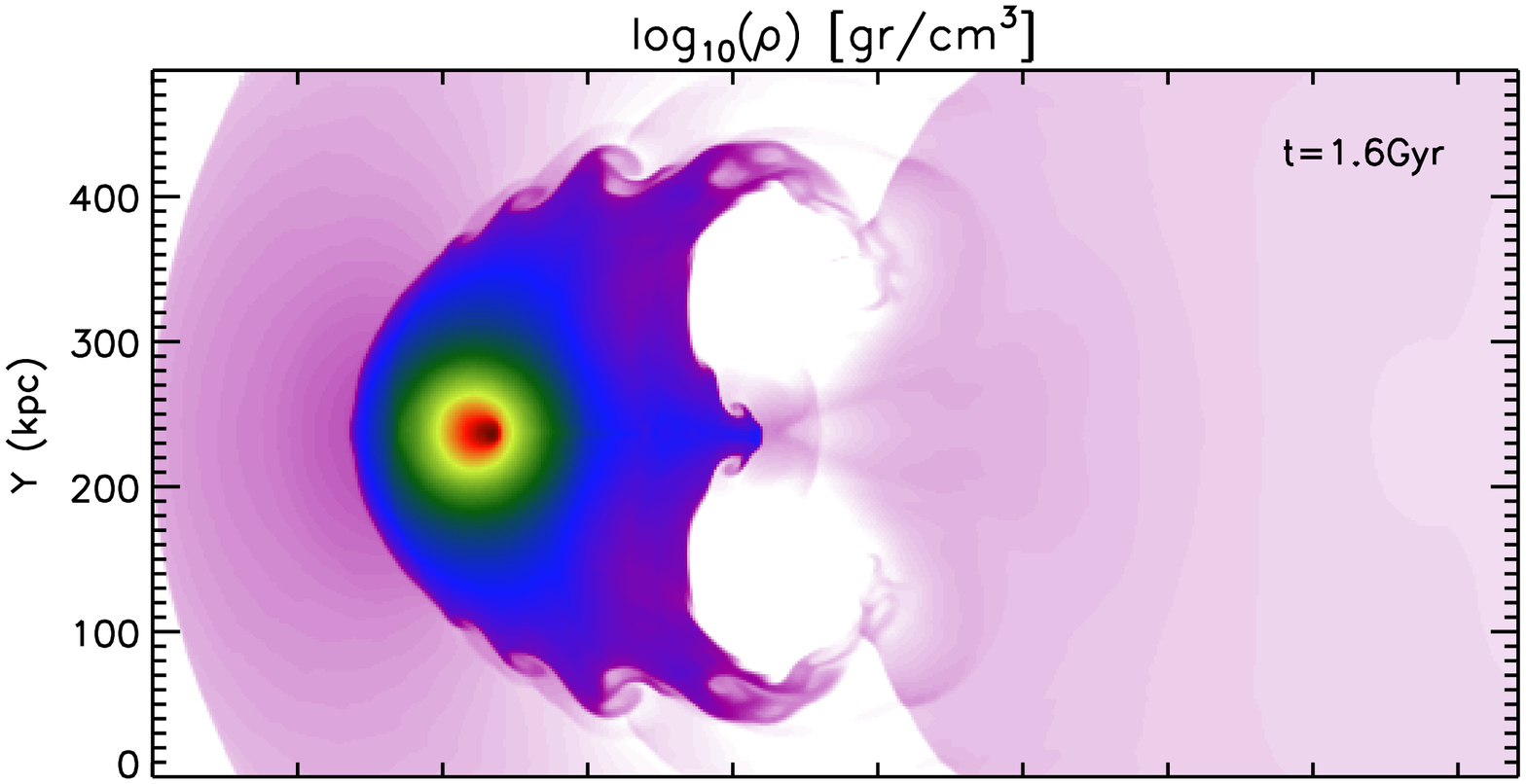} & 
\includegraphics[height=9.2cm,width=3.68cm,angle=90,trim=4.6cm 0cm 0.1cm 0cm, clip=true]{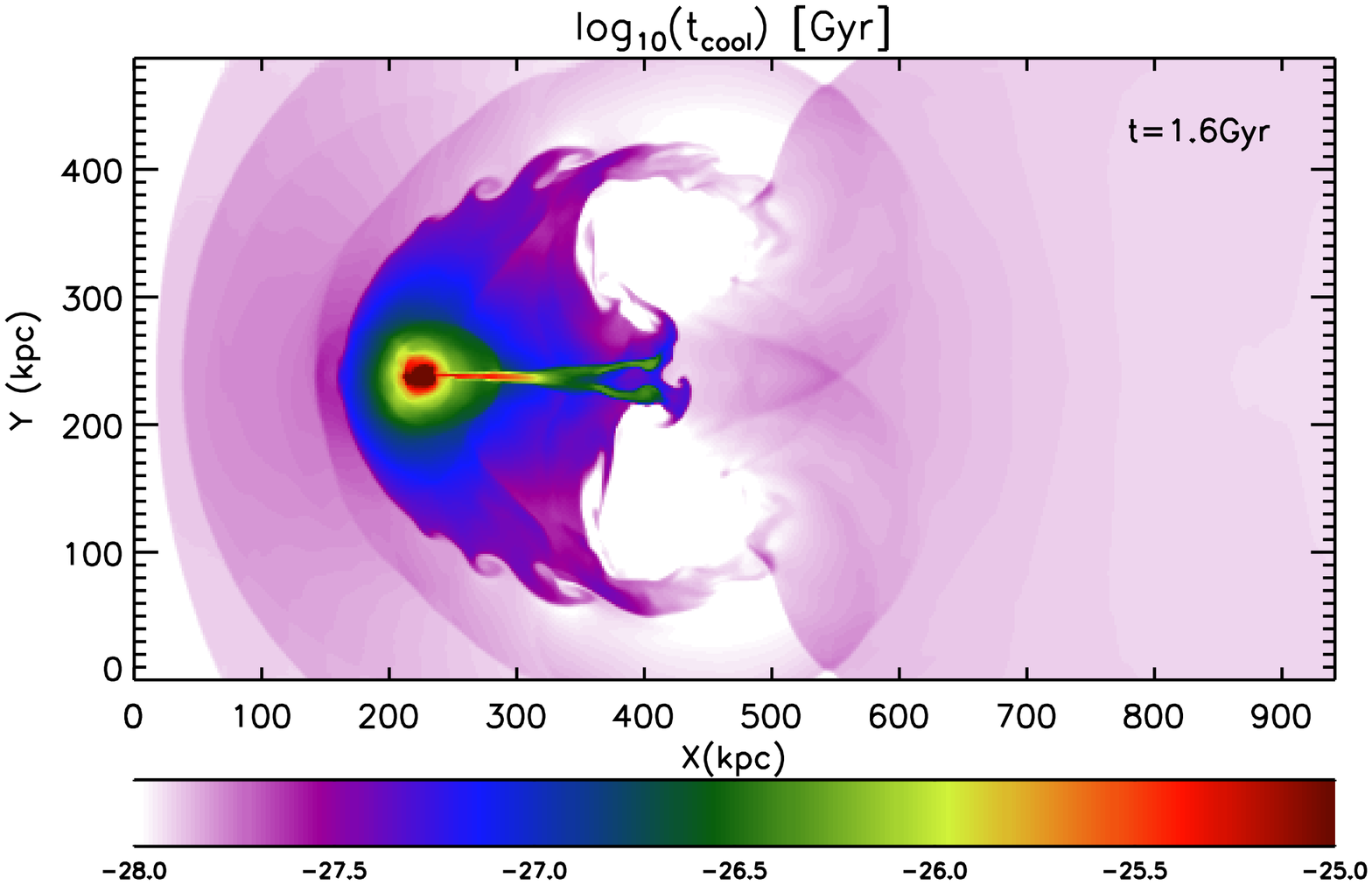} \\
\includegraphics[height=9.2cm,width=3.68cm,angle=90,trim=4.6cm 0cm 0.1cm 0cm, clip=true]{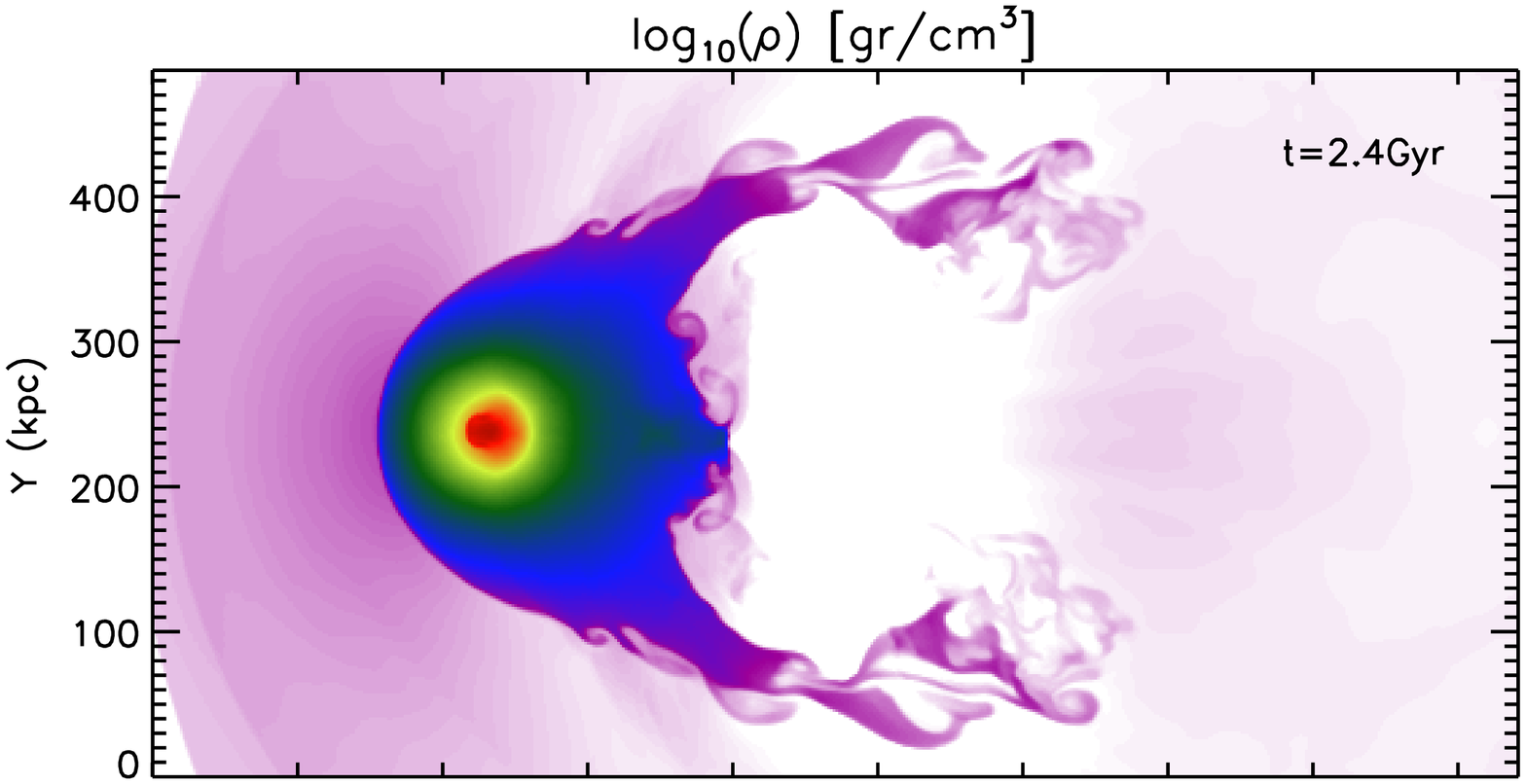} & 
\includegraphics[height=9.2cm,width=3.68cm,angle=90,trim=4.6cm 0cm 0.1cm 0cm, clip=true]{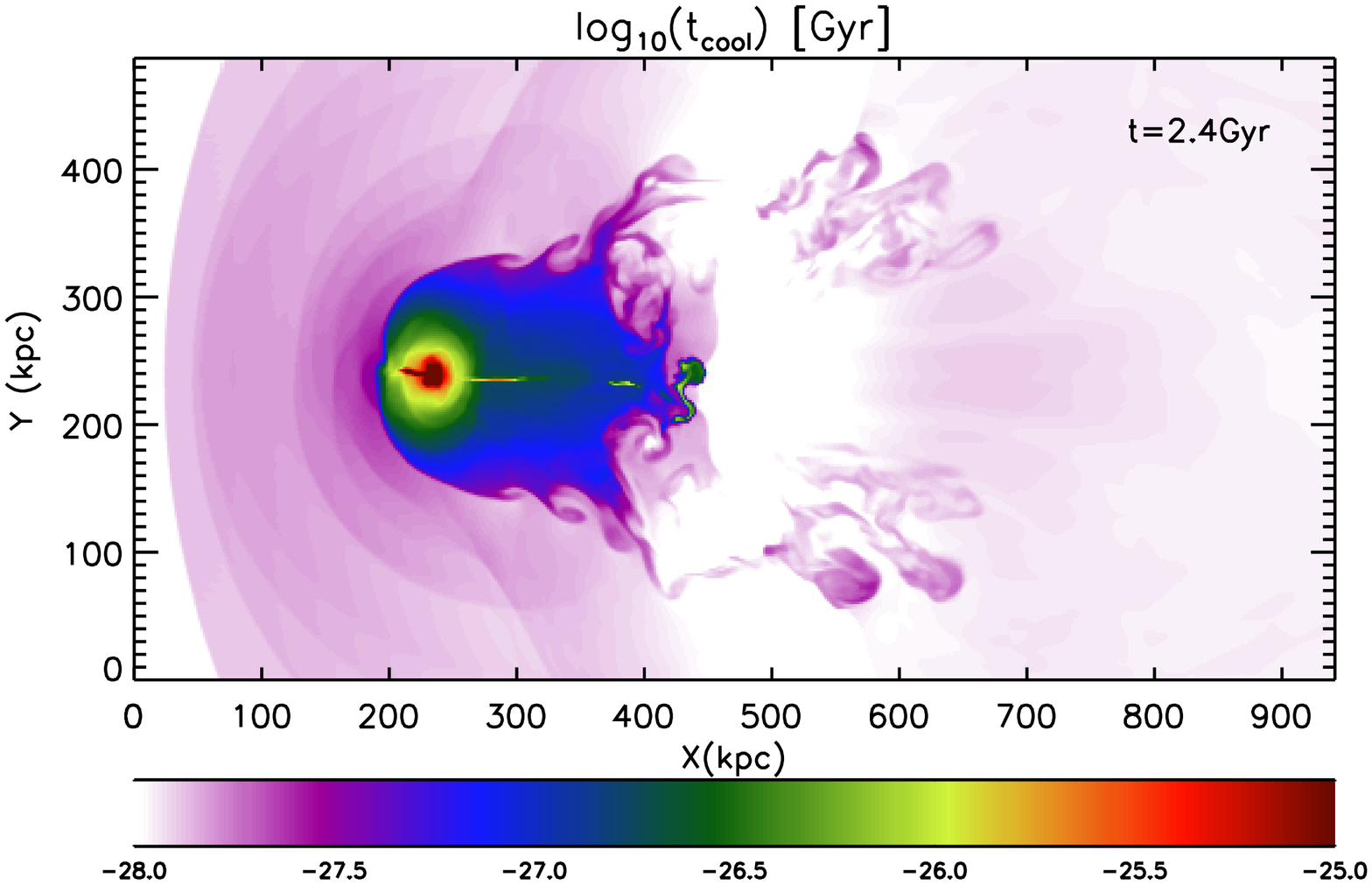} \\
\includegraphics[height=9.2cm,width=5.1cm,angle=90,trim=0.6cm 0cm 0.1cm 0cm, clip=true]{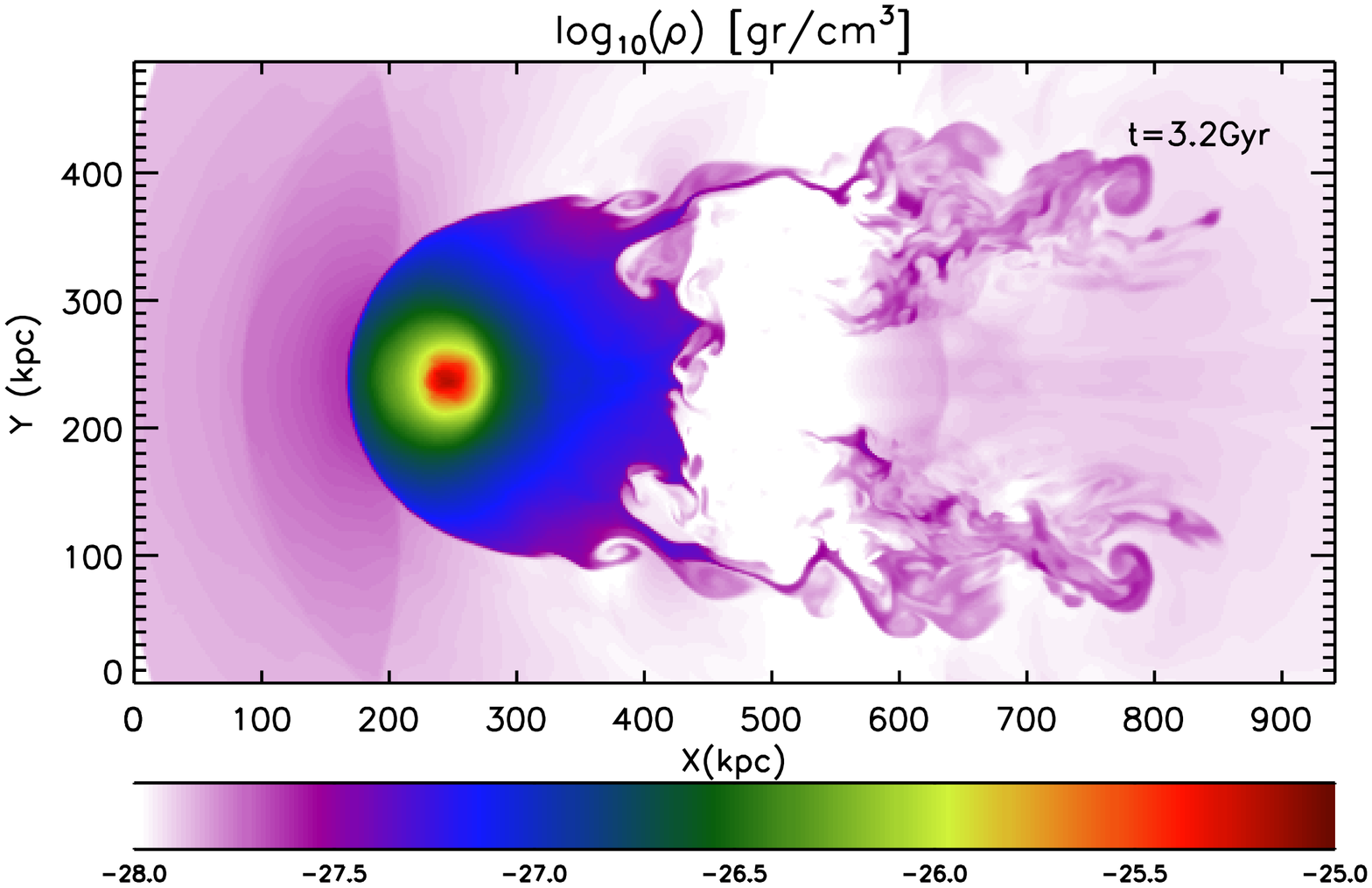} & 
\includegraphics[height=9.2cm,width=5.1cm,angle=90,trim=0.6cm 0cm 0.1cm 0cm, clip=true]{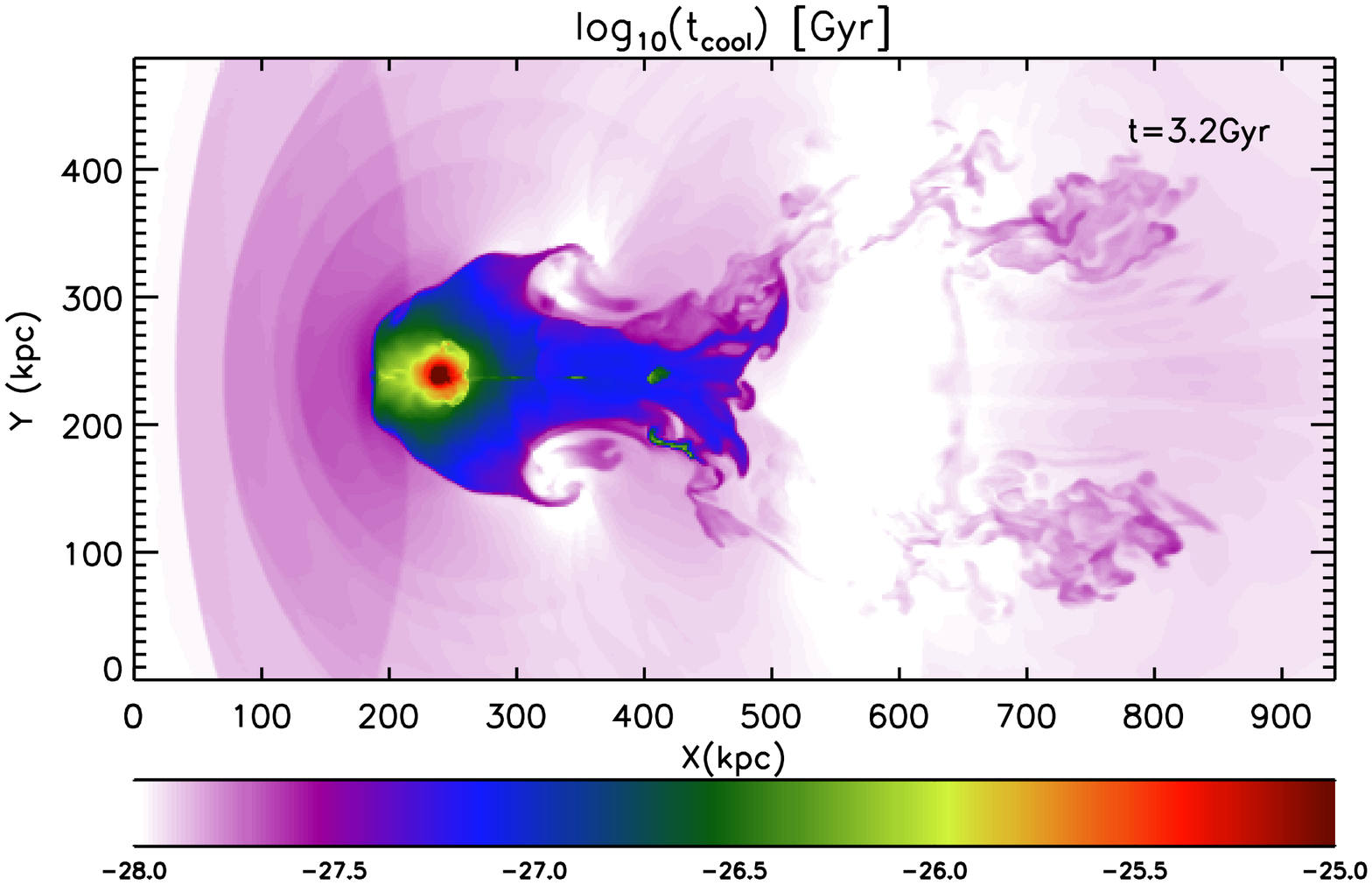} \\
\end{tabular}
\caption{Slices of gas density (logarithmic) in the XY plane through the middle of the simulation box for the non-radiative (left) 
and dissipative (right) simulations at $t=0.8,1.6,2.4$ and $3.2$~Gyr.}
\label{dens_temp} 
\end{figure*}

Gas is simulated by a hierarchy of cells using the AMR technique. The initial pressure in the cells is given by
the condition of hydrostatic equilibrium as described at the beginning of Section \ref{theo}. 
The velocities of the particles are set to zero. Finally the gas is set to have a mean molecular weight of $\mu^{-1}=1.71$.

In this way we generate an initial equilibrium configuration for a MW-size galaxy having the parameters given
in Table \ref{Table_1}. We have tested that this system remains in equilibrium for at least $\sim5$Gyr when evolved in isolation.
Nonetheless, two clear features appear in the density profile during this time: a small core ($\sim5$kpc) that develops in the centre
due to the limited resolution of the grid, and a small fraction of the dark matter particles ($\sim2\%$) expand to larger radii due
to the tapering of the density profile at the virial radius.

Finally, to simulate the ICM, the system is surrounded by gas with constant density ($10^{-28}$g~cm$^{-3}$) and temperature
($3.6\times10^{6}$K) moving relative to the satellite with a speed $v_{\rm sat}=V_{\rm v,group}=315$~km/s. The simulation box has a
length of $1.9$~Mpc with periodic boundary conditions\footnote{Periodic boundary conditions are necessary to use the parallel multigrid solver
for the Poisson equation in the version of FLASH we used.}. The resolution of the simulation uses 32 cells for the coarsest level with 6 
levels of refinement to give a 
maximum spatial resolution of $\sim1$~kpc. We use $50^3$ active particles to represent the dark matter halo of the MW-size
galaxy, and $100^3$ ($50^3$) tracer particles to sample the initial gas density distribution of the ICM (satellite).

We consider two simulations, one without radiative losses and one where the gas is allowed to cool radiatively. This allow us to
understand the impact of cooling in the formation of the turbulent wake: efficient
cooling in the corona of the satellite prior to the removal of the gas would increase its density and reduce its sound speed, and hence
suppress the KH instabilities, greatly reducing the amount of stripped gas. 
Central feedback is expected to reduce the cooling rate in the satellite, preventing a catastrophic collapse that would otherwise occur 
both in nature and in the simulations (this collapse also leads to prohibitive simulation time steps). 
To mimic the effects of feedback we multiply the cooling rate by a function of the distance to the center of the satellite: 1$-1/(1 + (r/r_{\rm c})^3)$
where $r_{\rm c}=0.15~r_{\rm v}$. In addition, we introduce a temperature floor of $10^5$K, below which the gas does not cool further.
We remark that this feedback does not affect the regions of interest, namely outside $\sim0.7r_{\rm v}$ (the stripping radius)
and that we do not intend to give a realistic feedback model but simply to attenuate the central overcooling.
We simulate a single gas species with a metallicity of log$_{10}(Z/Z_{\odot})=-1.5$ in both, the satellite and the ICM. 
Although in a realistic setting the gas is likely to have different metallicities according to its position relative to the center of the satellite,
and to the center of the low-mass group, we choose this value for simplicity and because it further help us reducing the central collapse of the
corona. We note that this value is roughly of the order of what is found in simulations of MW-size haloes
near the virial radius (see Fig.~9 of \citealt{voort_2011}).

\section{Results}\label{results}

Fig.~(\ref{dens_temp}) shows slices of the XY plane through the middle of the simulation box for the density of the gas 
for the non-radiative (left panels) and dissipative (right panels) cases at four output times $t=0.8,1.6,2.4$ and $3.2$~Gyr from top to bottom, 
respectively. The effect of RPS and the development of KH instabilities are clearly seen. In the top panels, we 
see how the shock has propagated through the satellite and is passing through its dense centre, whereas in the outskirts,
it has traversed most of the satellite. By $t\sim2.4$~Gyr, the shock has crossed the satellite completely and the gas
removed by RPS is already forming clear trailing wakes. This roughly agrees with our estimate of 
the RPS time scale to be of the order of 1.8~Gyr (see Table \ref{Table_2} and Appendix \ref{app_1}). 

In the dissipative simulation, the central collapse is already apparent after $t\sim1$~Gyr with the formation of a dense and cold
core. Although the artificial suppression of cooling prevents the central collapse from continuing further,
the contraction of the corona is large enough to significantly reduce the amount of stripped gas. However, the presence of the trailing wakes 
and the effect of turbulence are still evident. 
The linear, dense and cold feature to the right of the center of the satellite is caused by the asymmetry of the problem and the inclusion 
of a spherically symmetric cooling suppression. Although the shape of the central ``inner'' wake is affected by this feature, there is
only a minor impact on the extended ``external'' wakes that participate in most of the mixing with the ICM.

\subsection{Turbulence and gas mixing}

\begin{figure}
\centering
\includegraphics[height=8cm,width=8cm]{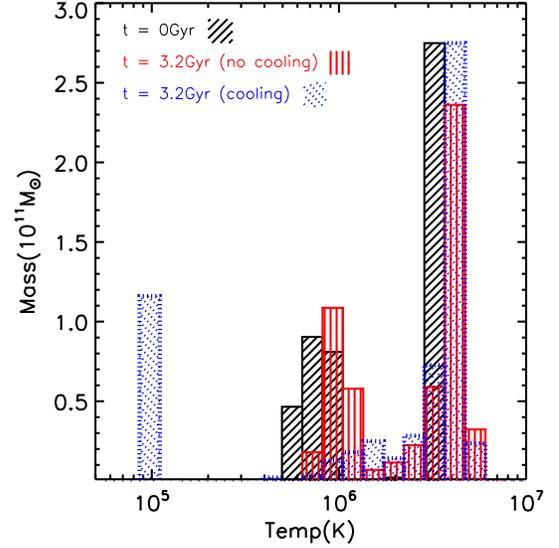}
\caption{Mass-weighted histogram of the temperature within a reduced volume of the simulation box, see text for details. 
The black histogram is for the initial simulation time, whereas the
red and blue histograms are for the final simulation time for the cases without and with radiative cooling, respectively.}
\label{temp_hist} 
\end{figure}

After $t\sim1~$Gyr, the stripped gas starts mixing efficiently with the ICM through the turbulence
generated by KH instabilities. From an initial look at the density and temperature of the gas at $t=3.2~$Gyr, it seems that the gas mixture 
has twice the density and nearly half the temperature of the ambient ICM. 
To give a first estimate of the amount of mass that makes the bulk of the mixed phase, we show in Fig.~(\ref{temp_hist}) a mass-weighted 
temperature histogram of all the mass contained within a box that extends above and below the slice shown in Fig.~(\ref{dens_temp}) 
with a thickness of $\sim490$~kpc. In this way, the volume covered by this box encloses all the mass that is 
being mixed, the satellite, and of course a fraction
of the ICM that is not involved in the mixing. The black histogram is for $t=0$~Gyr, while the blue and red are for $t=3.2$~Gyr for the
simulations with and without cooling, respectively. The initial gas distribution 
within the satellite is clearly seen around $10^6$~K as a ``cold'' phase, as well as the singled-value ``hot'' phase in the ICM to
the right. At the end of the non-radiative simulation, the gas in the satellite has been shock-heated while the ICM has been adiabatically 
compressed and expanded (see below). For the dissipative simulation, the centre of the satellite has partially collapsed, forming a cold core
with a temperature given by the imposed floor of $10^5$~K. The rest of the gas however seems to be distributed in a similar way in both simulations; 
in particular, the formation of a ``warm'' phase at $T\sim2\times10^6$~K is apparent in the figure for both cases. We show below
that this phase is actually formed by the mixture of the two phases and not just by the gas stripped from the satellite. 

\begin{figure*}
\begin{tabular}{|@{}l@{}|@{}l@{}|}
\includegraphics[height=4.5cm,width=9cm, trim=-0.43cm 1.5cm 0.82cm 2.5cm, clip=true]{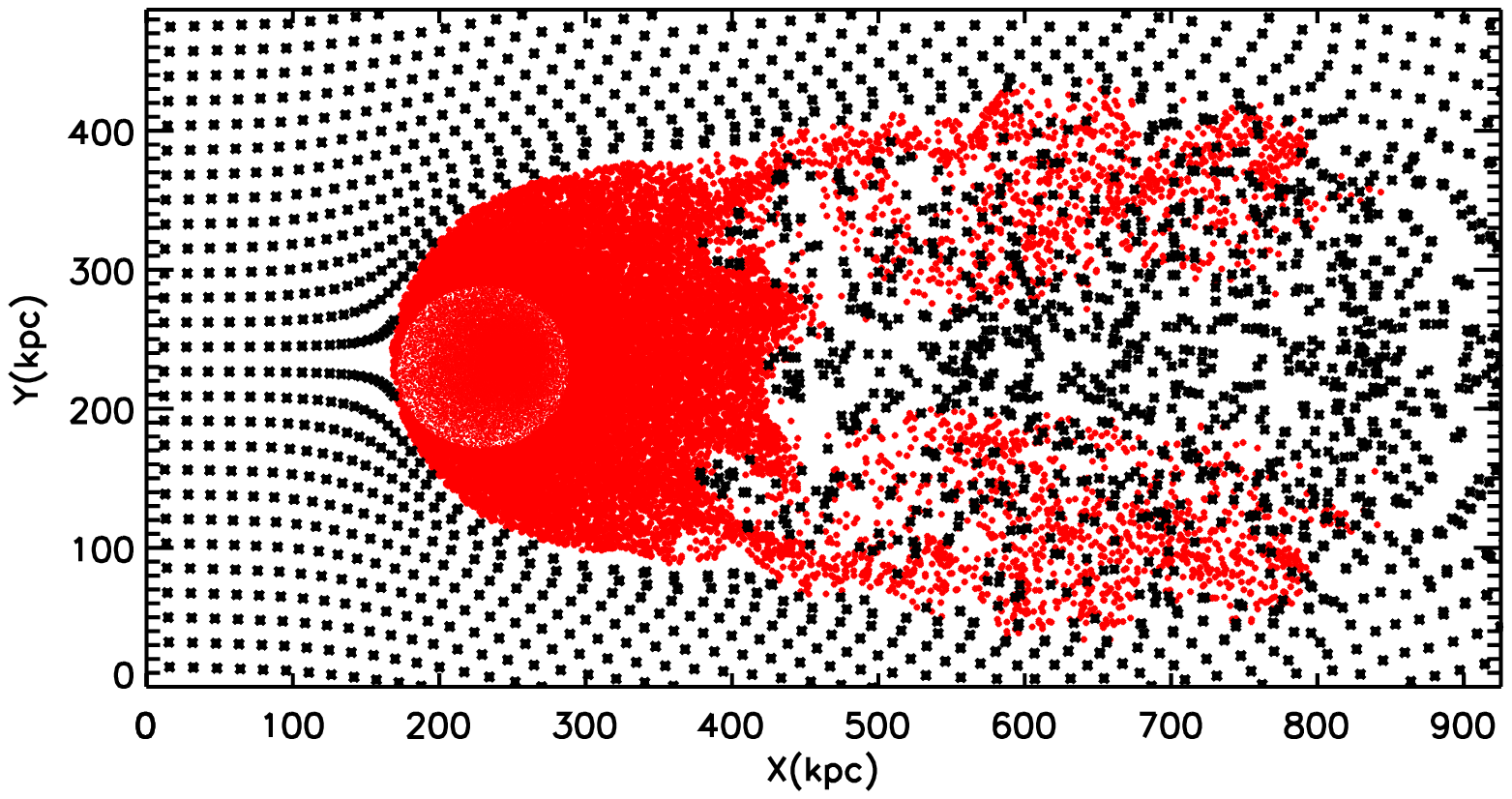} &
\includegraphics[height=4.5cm,width=9cm, trim=-0.43cm 1.5cm 0.82cm 2.5cm, clip=true]{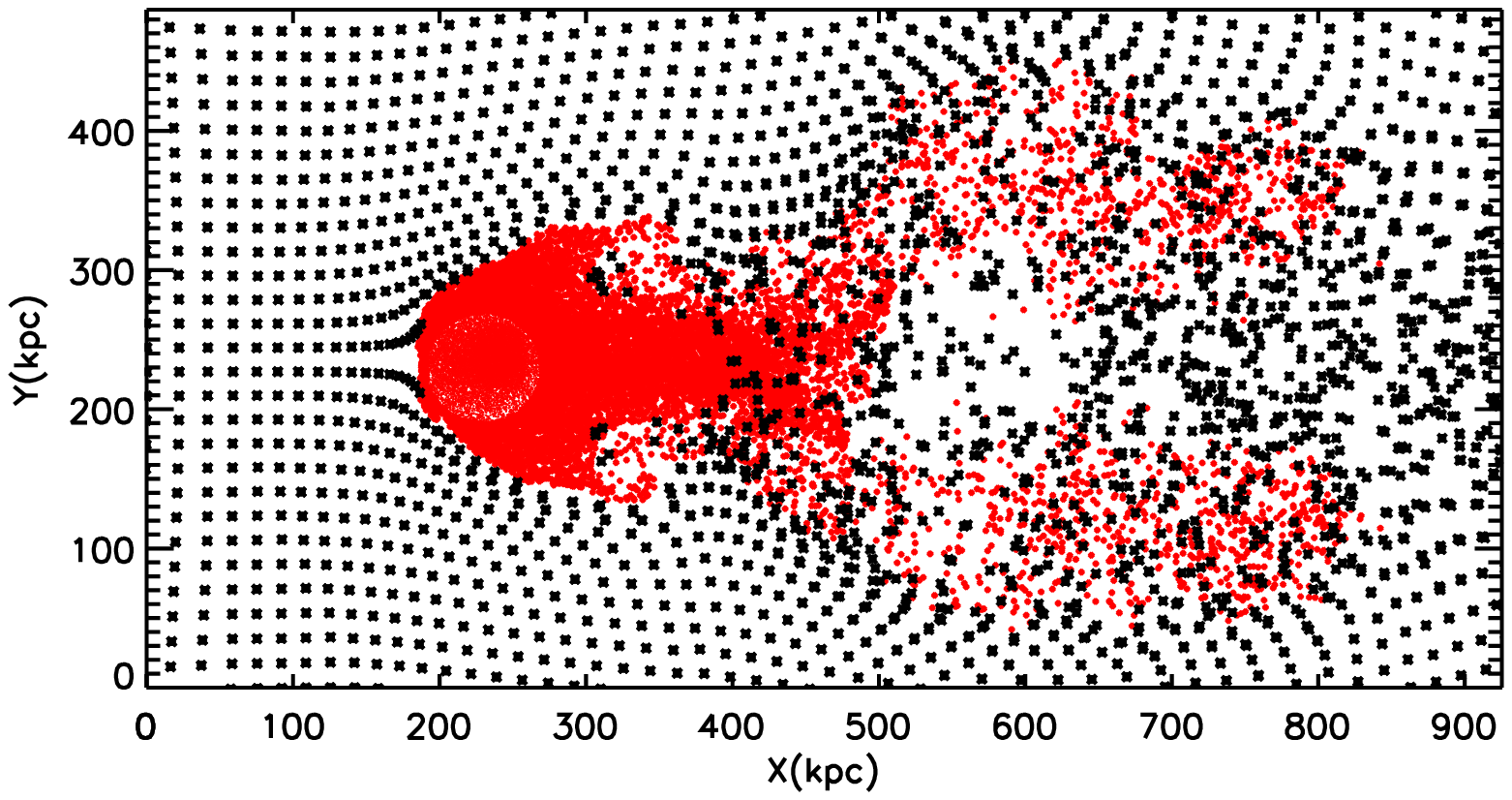} \\
\includegraphics[height=9.45cm,width=5.1cm,angle=90,trim=0.6cm 0cm 0.1cm -0.5cm, clip=true]{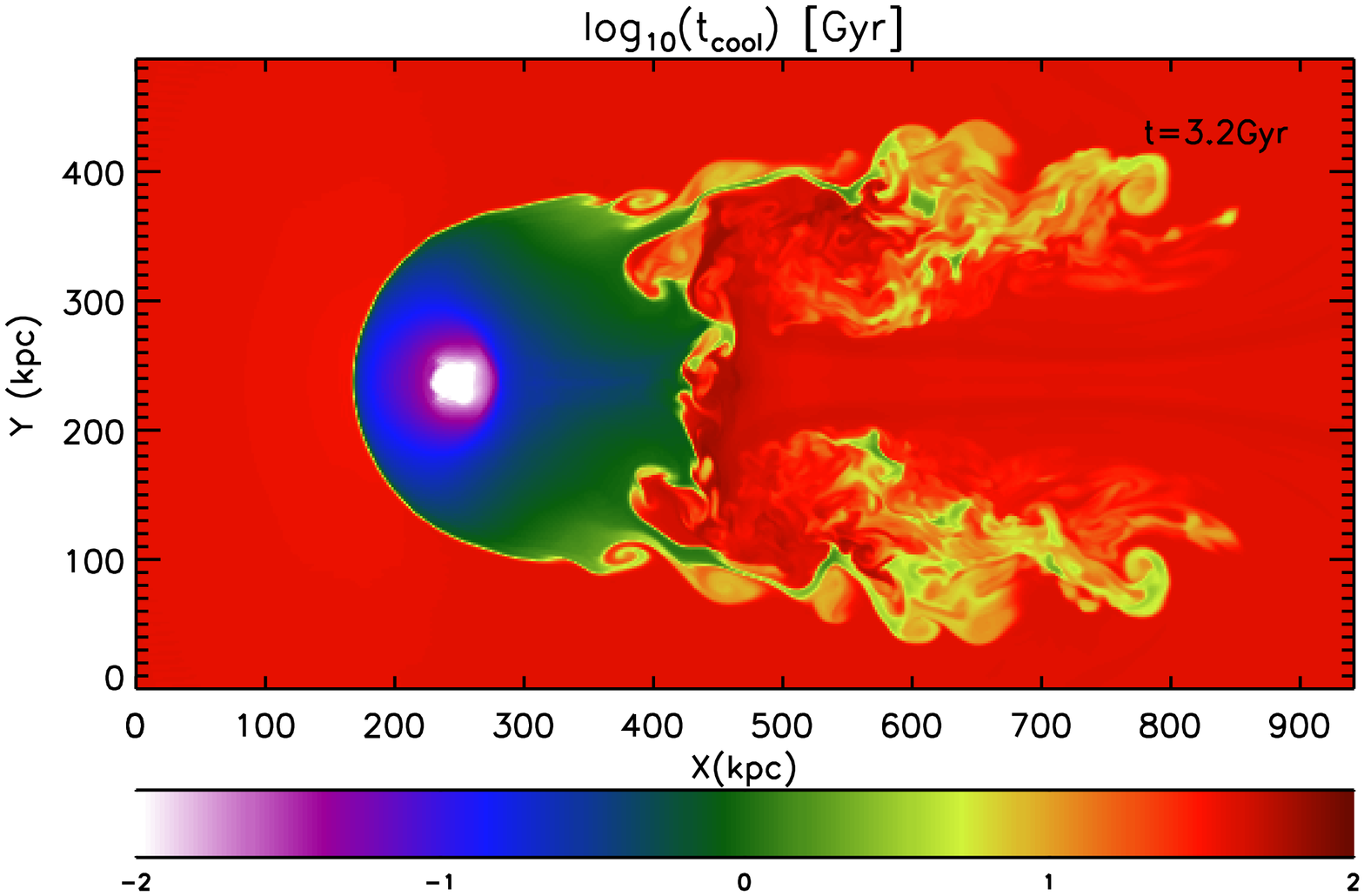} &
\includegraphics[height=9.55cm,width=5.1cm,angle=90,trim=0.6cm 0cm 0.1cm -0.5cm, clip=true]{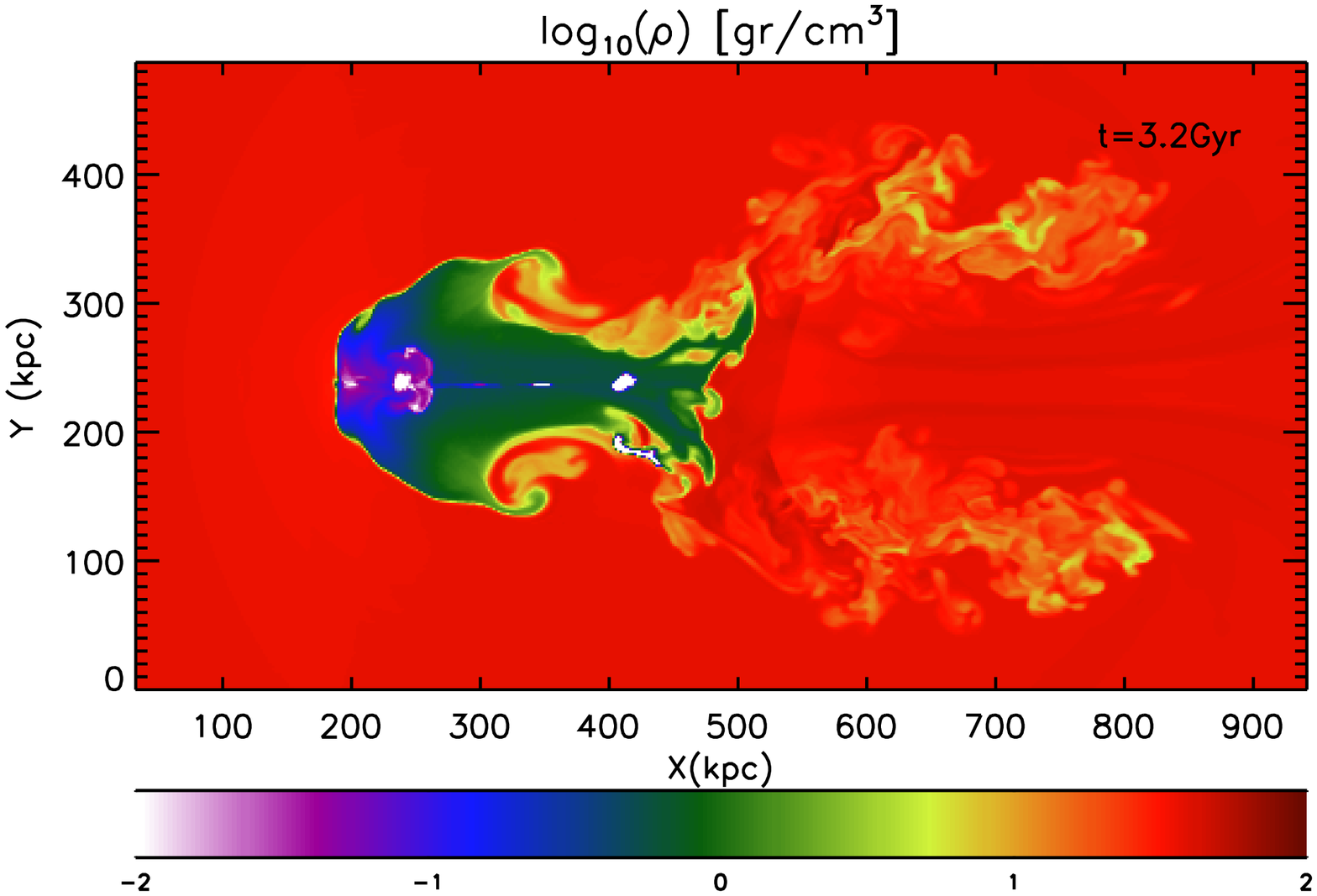} \\
\end{tabular}
\caption{{\it Upper panels:} Distribution of tracer particles after $t=3.2$~Gyr for a slab through the middle of the simulation box with a 
thickness of 50~kpc; red and black for the satellite and ICM, respectively. {\it Bottom panels:} Slice in the XY plane through the middle of
the simulation box at $t=3.2$~Gyr showing the values of the cooling time (logarithmic) assuming log$_{10}(Z/Z_{\odot})=-0.5$. 
The panels on the left (right) correspond to the
simulation without (with) radiative cooling.}
\label{mix_cool_fig} 
\end{figure*}

The upper panels of Fig.~(\ref{mix_cool_fig}) give an idea of the degree of mixing between the fluids at the end of the non-radiative 
(left panel) and dissipative (right panel) simulations. They show the distribution of tracer particles that were set at $t=0$ to sample the density 
distribution in the satellite (red) and in the ICM (black). In the figures,
a projection of this distribution at $t=3.2$~Gyr is shown for a slab with a thickness of 50kpc\footnote{The tracer particles sample
the gas distribution with lower resolution than the mesh cells, particularly for the ICM where the initial inter-particle separation is $\sim15$~kpc.}.
The objective of this figure is to show that the two fluids
originally separated are thoroughly mixed by the turbulent motions in the plumes. The degree of mixing is similar in both simulations.

The bottom panels of Fig.~(\ref{mix_cool_fig}) are a slice through the middle of the simulation box showing the cooling time, constructed using the
temperature and density of the gas at $t=3.2~$Gyr and assuming a single metallicity, log$_{10}(Z/Z_{\odot})=-0.5$, representative of 
the intragroup gas. In reality, the cooling time of the gas mixture would depend on the mass-weighted metallicities of the species being mixed. Since
these are uncertain, and for simplicity, we have chosen a constant metallicity, but of course, the absolute value of the cooling
times reported in this work are metallicity-dependent. We see that in the trailing wakes where the gas mixes 
more efficiently, the cooling times have decreased by up to an order of magnitude compared to the original cooling time in the ICM. 

\begin{figure*}
\begin{tabular}{|@{}l@{}|@{}l@{}|}
\includegraphics[height=9.1cm,width=9.1cm]{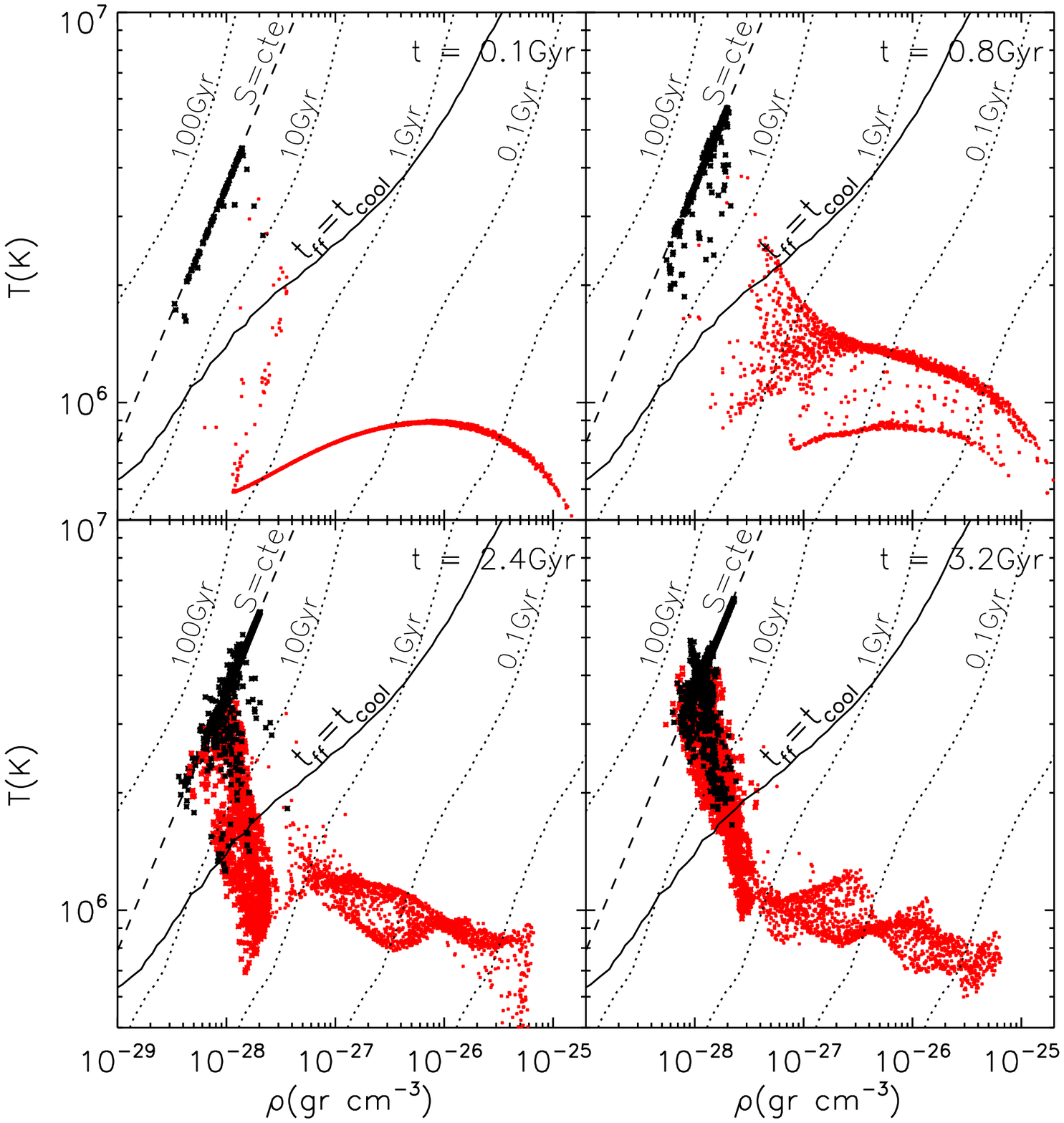} &
\includegraphics[height=9.1cm,width=9.1cm]{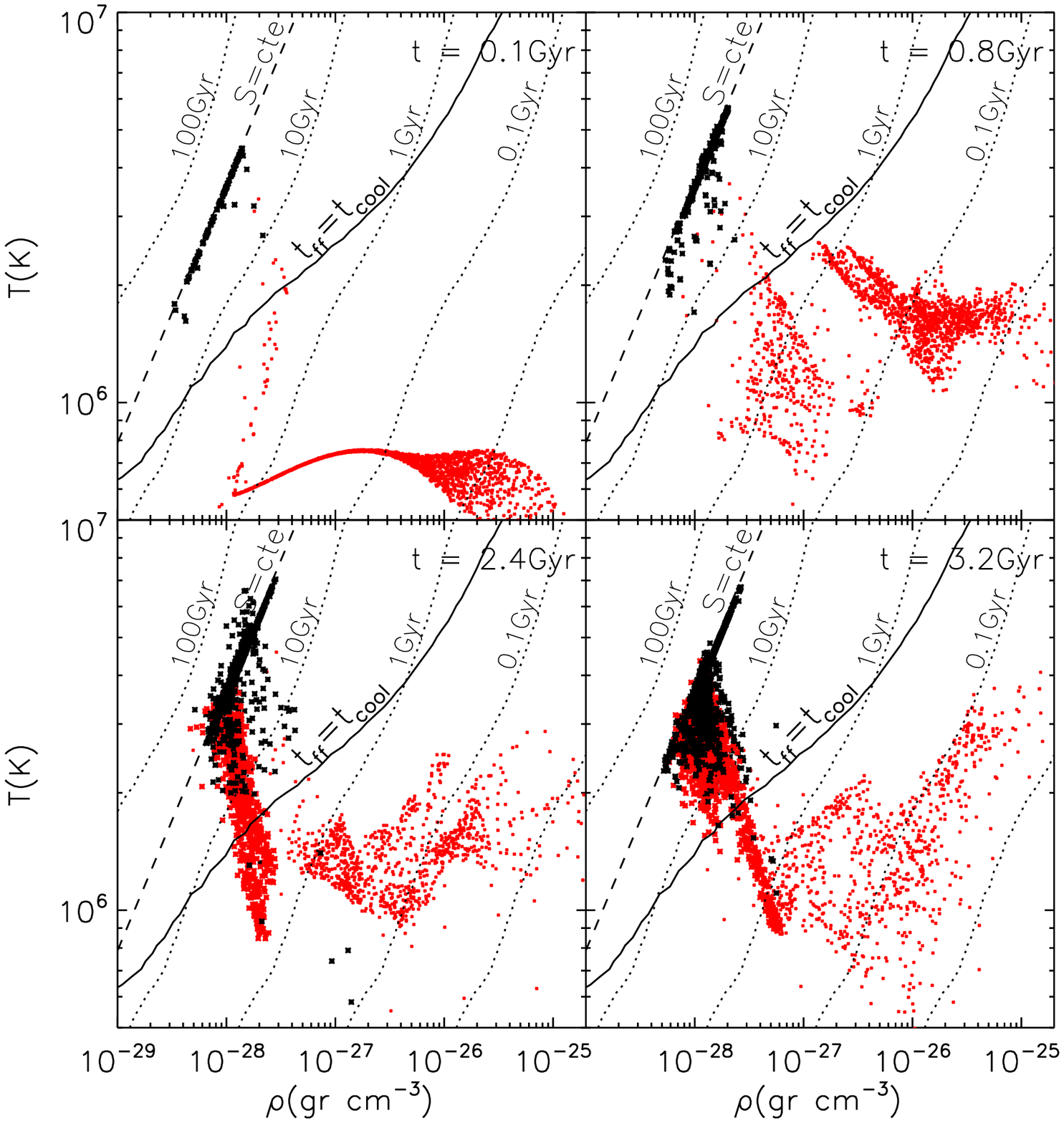} \\
\end{tabular}
\caption{Temperature-density diagram for the tracer particles contained within the volume
that appears in the upper panel of Fig.~\ref{mix_cool_fig}. Only a random subsample of the full tracer
population within this region is plotted (except in the bottom panels for the region around the wakes, $X>500$~kpc in Fig.~\ref{dens_temp},
where the tracers are shown with bigger symbols. The red (black) points are for the satellite (ICM). The simulation times
shown are $t=0.1,0.8,2.4$ and $3.2$~Gyr. The dotted lines are lines of constant cooling time (assuming log$_{10}(Z/Z_{\odot})=-0.5$) and 
the black solid line shows the locus of
points where $t_{\rm ff}=t_{\rm cool}$. The dashed lines mark an example of a path with constant entropy. 
The set of panels to the left (right) corresponds to the simulation without (with) radiative cooling.}
\label{phase_diagram} 
\end{figure*}

Fig.~(\ref{phase_diagram}) shows the position of the tracer particles (in red and black for the satellite and ICM, respectively), 
within the slab shown in the upper panel of Fig.~(\ref{mix_cool_fig}) (i.e., 
within a box of volume $942\times487\times50$~kpc$^3$) in a temperature-density diagram at different simulation times: $t=0.1,0.8,2.4$
and $3.2$~Gyr clockwise starting from the upper left of each set of panels. The set to the left (right) corresponds to the simulation without
(with) radiative cooling. The dotted lines are lines of constant cooling time
from 100~Gyr to 0.1~Gyr going from left to right. The thick black solid line shows the locus of points where the cooling time equals the local
free-fall time ($t_{\rm ff}=t_{\rm cool}$). Finally, the dashed line is a reference marking processes that conserve entropy (defined here
as $S\propto T/\rho^{2/3}$). 

Since the number of tracer particles in the satellite is high (particularly in the center of the satellite), we have randomly selected a smaller 
sample to plot in Fig.~(\ref{phase_diagram}), except in the region around the wakes for the last two time outputs, bottom panels, where 
the tracers are shown with bigger symbols. We have verified 
that this random subsample traces the distribution of the full sample and remark that this was done simply for plotting purposes. 
This region around the wakes is the one corresponding to $X>500$~kpc in Fig.~(\ref{dens_temp}) (see section \ref{fuel}). 

Initially, the ICM gas represents a single point in this diagram ($\rho=10^{-28}$g~cm$^{-3}$, $T=3.6\times10^6$~K) 
and the gas in the satellite is distributed according to the equilibrium configuration. In the
first 100~Myr, the gas in the ICM immediately surrounding the satellite compresses upstream and expands downstream, conserving its entropy. The
gas on the satellite is compressed and shocked upstream; the gas loosely bound to the satellite begins to be removed. 
As the shock traverses
the satellite, it heats and compresses the gas within it, removing material from the outskirts while gravity tends to restore the 
original gas distribution; the distribution within the inner regions will remain distorted though, since the shock causes an irreversible 
change in the gas. 

In the dissipative simulation, the gas that remains within the satellite is additionally subjected 
to the central collapse, which is suppressed in the innermost regions by the feedback-like
mechanism we have introduced. This creates a gas distribution that is highly dispersed in the phase diagram (the coldest and densest material is not
explicitly shown in Fig.~\ref{phase_diagram} since we concentrate on the regions where mixing happens). On the other hand, the behaviour 
of the gas in the outskirts of the satellite, and outside in the ICM, is qualitatively similar to the one in the non-radiative simulation.
In particular, most of the ICM evolves along a path of constant entropy (expanding and contracting driven by the relative 
motions of the fluids, and the pressure support and gravity of the satellite). At $t\sim1$~Gyr, a fraction of the ICM is clearly mixing with the 
stripped gas increasing its density and decreasing its temperature. By $t=3.2$~Gyr, a mixed phase of gas is apparent in 
the figure with $T\sim2\times10^6$~K and $\rho\sim2\times10^{-28}$g~cm$^{-3}$. This confirms the initial visual impression given 
by the bottom panels of Fig.~(\ref{dens_temp}), and also proves that the ``warm'' phase shown in Fig.~(\ref{temp_hist}) is
a mixed phase formed by gas from the satellite and the ICM. 

By looking at the lines of constant cooling time, we see that the mixed phase reaches values of $t_{\rm cool}\sim5$~Gyr, considerably lower
than the initial cooling time of $\sim40$~Gyr. We also see that by the end of the simulation, the mixed phase is approaching the locus of points
$t_{\rm ff}=t_{\rm cool}$ that separates the gas that can cool efficiently from the one that cannot. 

\subsection{Properties of the wakes}\label{fuel}

We define the wakes as the trailing structures that are clearly distinguished in Fig.~(\ref{dens_temp}) at $X>500$~kpc. By doing so, we exclude
the material that is not bound to the satellite but is still attached to it. This gas will be stripped in the future and added to the 
trailing wakes, but since this gas is clearly not participating in the mixing we omit it from the analysis. We note that 
although this definition is arbitrary, the properties of the wakes at a given time serve as an
approximate reference for the properties at other times.
With such definition, we find that at $t=3.2$~Gyr roughly $50\%$, $55\%$ of the mass that is no longer bound to the satellite is
in the trailing wakes in the simulations with and without radiative cooling, respectively.
In the former the amount of stripped gas is $\sim7\times10^9$M$_\odot$, a factor of 3 smaller than in the latter. 

The gas in the ICM mixes with the gas from the satellite and becomes part of the trailing wakes. Although this mixing occurs in a wide region,
we can arbitrarily define the ``warm phase'' as the region in the wakes where the gas (both from the ICM and from the satellite) has an entropy which
is at least 50\% lower than the initial entropy of the ICM. Gas with higher entropy has either not been thoroughly mixed yet and/or contains the 
highest entropy gas from the satellite. 
Using this definition, we can narrow down the median values of the temperature and density of the gas that is more likely to cool afterwards
and estimate its total mass after $3.2$~Gyr of the evolution for the simulation without (with) radiative cooling: 
$\left<T_{\rm mix}\right>\sim2(2.1)\times10^6$K, 
$\left<\rho_{\rm mix}\right>=1.8(1.8)\times10^{-28}$g~cm$^{-3}$, and M$_{\rm mix}=7.6(1.4)\times10^9$M$_{\odot}$ ($\sim20(28)\%$ coming from the ICM). 
These values imply $t_{\rm cool}\sim7(8)$Gyr, 
a factor of $\sim6(5)$ less than that of the ambient ICM. 
In the non-radiative simulation, most of the gas in the wakes that came from the satellite belongs to the warm phase (59\%), while in
the dissipative simulation is a minority (27\%). This is simply explained by the smaller amount of stripped gas that participates
in the mixing in the dissipative simulation, which also explains why the warm phase has a longer
cooling time in this case. 

\section{Discussion}\label{sec_discussion}

The relatively weak impact of radiative cooling in the formation and initial properties of the wakes (with the main
difference being a reduction of the stripped gas) can be understood by noting the following.
The gas in the satellite that participates in the mixing is given, at the most, by the gas within a shell external to the stripping radius.
Even if all the gas within this radius were to collapse immediately due to radiative losses, the outer shell would collapse 
in a free-fall time, which is $\gtrsim1$ Gyr; this contraction would happen adiabatically since the gas in the shell has still 
quite a long cooling time. An adiabatic process preserves the cooling time at the temperatures and densities of the collapsing shell
(note how the lines of constant entropy are roughly parallel to the lines of constant cooling time in Fig.~\ref{phase_diagram}). 
This is because the cooling function goes as $\Lambda_{\rm cool}\propto T^{-\alpha}$, with 
$\alpha\sim0.5$ for log$_{10}(Z/Z_{\odot})\sim-0.5$, in the temperature range between $10^5$K and $\sim5\times10^6$K. 
Thus, the gas in the shell, that is eventually stripped, has enough time to mix with the ICM forming a wake in a qualitatively similar way as in the
non-radiative simulation.

In detail, the properties of the wakes depend on the evolutionary stage of the interaction and, of course, on the
metallicity, temperature and density of the gas being stripped. The latter depend on the assumed radial profiles and have important consequences 
in our results. Although we cannot check directly from observations if these profiles are adequate, we
found that the the properties of the gas near the outskirts of the corona are roughly the same as the ones reported in \citet{Crain_2010}
(based on a large sample of MW-size galaxies extracted from a cosmological simulation). The mass of
stripped gas not only depends on the balance of radiative cooling and central feedback
(which determines how fast the central collapse relative to the stripping time scale is) but also in the
uncertain amount of hot gas that is in the corona. In this work we have used the largest possible value
set by the universal baryon fraction. However, e.g. \citet{Anderson_2010,Anderson_2011} found that the actual amount of mass in
the hot coronae is considerably less than naively expected from subtracting the universal baryon fraction from the amount of baryons that 
is locked in stars and cold gas. A lower gas fraction would lower the mass of stripped gas. 
Because of all this, our predictions cannot be made in terms of the absolute values of the properties of the gas that mixes, 
but rather in terms of the mixed gas fraction relative to the stripped gas,
and its cooling efficiency relative to the temperature and density of the removed gas.

Our non-radiative simulation indicates that the analytic model we introduced in Section \ref{theo} gives reasonable
estimates. In particular, once we apply the model to
the conditions in the simulation, we find that it is able to reproduce some of the global properties of the wakes (see below)
once we adjust its parameters and add two additional components to it: i) The stripped gas 
is shock-heated to temperatures in excess of the virial temperature of the satellite before it mixes with the ICM. We can account
for this effect by calculating the Rankine-Hugoniot (RH) temperature jump conditions of the shock (leading to a shock-heated temperature
$\sim1.7\times10^6$~K); ii) The gas in the wakes is compressed by a factor $\delta_{\rm mix}$
relative to the density of the ICM. With these considerations we find that by taking: $\alpha_{\rm rem}\sim 1$, $\beta\sim3$, and 
$\delta_{\rm mix}\sim2$ we are able to reproduce the amount of stripped gas, the median temperature in the mixed
phase and its cooling time. Although comparing the analytical model to the dissipative simulation is in principle more complicated due to
the contraction of the corona, it is possible to use the current model without modifications if we assume that the contraction occurs rapidly
relative to the gas removal processes. Then, the results of the dissipative simulation can be matched  by setting
$\alpha_{\rm rem}=1/3$ and $\beta=1.5$. The first of these parameters is reduced, relative to the non-radiative case, to accommodate a smaller
amount of stripped gas, while the second is reduced since the corona is more compact.

Perhaps the most important limitation of the model arises from the complex nature of the mixing process which
occurs in the turbulent eddies in layers of different size. This implies that
defining an overall effective value for $\beta$ is unrealistic since there is a broad distribution of cross sectional areas. 
In the simulations, the mixing happens along the annulus formed by the wake, which has a cross sectional area
that broadens if the relative 
velocity of the satellite decreases. Although the gravitational focusing effect is stronger in this case, the gas in the turbulent eddies is 
confined to a region which is more extended than the case of larger velocity
since the ram pressure of the surrounding ICM is smaller (we have corroborated this by running an identical simulation but with 
$v_{\rm sat}=150$km/s). This suggests that $\beta$ increases (decreases) for lower (higher) velocities,
contrary to expectations. 

Despite its simplicity, the analytic model could
be used for instance in semi-analytic models (SAMs) of galaxy formation (for a review see e.g. \citealt{Baugh_2006}) to account for the
removal processes of the coronal gas, which are typically severely simplified by assuming that this gas is removed instantaneously
having no further importance.

There are specific conditions one can think of that would enhance the mechanism proposed in this work. For instance, 
if the infalling satellite has lower mass and hence a colder corona, the temperature in the mixture would be lower than
the case we analysed, and gas removal would be more efficient. A lower infall velocity would create a weaker shock and
thus a reduction on the temperature of the shock-heated gas as it is stripped from the satellite\footnote{The typical satellite 
infall velocities are of the order of the virial velocity of the group, lower infall
velocities are nevertheless possible \citep[e.g.][]{Wetzel_2011}. For example, if $v_{\rm sat}=150$~km/s, then
the temperature ratio of the post- to pre-shock gas according to the RH temperature jump conditions is reduced from
2.8, in the case when $v_{\rm sat}=315$~km/s, to 1.6.}. 
On the other hand, the strength and time scale of gas removal decrease 
making the mixing process less efficient.  

Although more speculative, we also expect that the mechanism would be more important as the satellite spirals towards the centre of 
the dense stratified medium of the host. We can crudely estimate what would
happen in these circumstances using our analytic model. 
Although the process is of course gradual, let's assume that by the time the satellite is at a distance of $0.5r_{\rm v,host}$
from the centre of the host, it has lost an amount of dark matter and gas as predicted by the analytic estimates at the virial radius, i.e.,
the satellite arrives at this orbital distance with a mass profile ``clipped'' at a tidal stripping radius $r_{\rm t}=0.65r_{\rm v}$
(see Table \ref{Table_2}). If the specific angular momentum of the satellite was conserved along its orbit, then its velocity would be
twice the velocity at the virial radius. Dynamical friction however, decelerates the satellite; for simplicity we assume
it to be $\sim1.5v_{\rm sat}$\footnote{We ran a dark-matter only simulation of a MW-size halo merging with a
$10^{13}$M$_{\odot}$ group. The initial infall velocity of the satellite, with its radial and tangential components, was set to the average
value as given in \citet{Wetzel_2011}. We found that the satellite reaches an orbital distance of $0.5r_{\rm v,host}$ 
with a speed of $\sim1.6v_{\rm sat}$ 
with approximately 70\% of its original dark matter mass. This simulation confirms the orbital evolution we have assumed here.}. 
The temperature in the corona of the satellite is that left by the remnant of the
initial shock, which is given by the RH jump conditions of the virial shock. The temperature
and density of the ICM at this position are $3.5\times10^6$K and $6\times10^{-28}$g~cm$^{-3}$, respectively. 
Under these conditions, we applied our analytic method and find that the 
remaining corona would be removed in a time scale roughly given 
by the sound crossing time $\lesssim1$~Gyr, and the gas in the mixture would have a cooling time of only $1.5$~Gyr (a factor of $\sim4$ less than 
the ambient ICM). 

It is worth mentioning that recent hydrodynamical simulations of galaxy formation performed with the code AREPO 
\citep[][a hybrid method combining some of the advantages of Lagrangian and Eulerian codes and avoiding their most 
important weaknesses]{AREPO_2010} 
have reported
significantly higher star formation rates for central galaxies in massive haloes than the ones found, under the same conditions, in an SPH
formulation \citep{Keres_2011}. The authors advocate that the difference is partly due to a more efficient mixing in AREPO, 
caused by a larger removal of low entropy material from infalling satellites \citep{Mark_2011}. Fig. 4 of \citet{Keres_2011} shows that for 
a $\sim10^{13}$M$_{\rm \odot}$ halo at the present day, this difference is a factor of $\sim2$: AREPO predicts approximately 
$3\times10^{11}$M$_{\odot}$ of additional stars. It is unlikely that the majority of this extra mass can be accounted for by the coronae 
mixing studied in this paper, however it might be 
that a significant portion of the additional star formation material comes from a similar mixing process between the ICM and the cold 
gaseous disc of the merging galaxy. We believe it would be interesting to explore the relative 
contribution of both of these processes.

\section{Conclusions}\label{conclusions}

The presence of an extended corona of hot gas within clusters, groups and massive galaxies is a generic prediction of the
current theory of galaxy formation. 
Since such coronae are a large gas reservoir that can potentially cool down and form stars, it is of great relevance
to study their evolution and in particular their interaction during
galaxy mergers which are ubiquitous in the CDM paradigm.

Most studies of galaxy mergers have concentrated on the interaction of the cold
components (stars and gaseous discs), relegating the hot coronae to a secondary role essentially governed by ram 
pressure stripping. In the case of MW-size galaxies merging into groups, the high temperature and low 
density of the ambient ICM at the outskirts of the groups are commonly assumed to imply that the stripped gas stays irrelevant 
for subsequent cooling and star formation. This
assumption, however, ignores the fact that Kelvin-Helmholtz instabilities produced by the velocity shear also remove
coronal gas and efficiently mix it with the ICM through turbulent eddies. 
Since the former is colder than the latter, the mixture could have a shorter cooling time than the ambient intragroup gas. 
These instabilities are typically suppressed in the majority of galaxy formation simulations that have been done to date 
since most of them were carried out using the SPH formulation \citep{Agertz_2007}.

In this paper, we use analytic methods and hydrodynamical simulations (with and without radiative cooling) to
study the gas removal processes that act during the interaction a of a spherical MW-size corona moving through a homogeneous 
ICM wind representative of the coronal gas at the virial radius of a low-mass group. The main conclusions of our work are the following:
\begin{itemize}
\item Our simulations show that a relevant fraction of the removed coronal gas
mixes with the ICM in trailing wakes creating a ``warm phase'' (i.e. with at least 50\% less entropy
than the ambient ICM) with a cooling time which 
is almost an order of magnitude lower than the cooling time of the ambient intragroup gas. This reduction is a result of 
a lower temperature in the gas mixture driven by the colder coronal gas from the satellite, as anticipated, but also due 
to an enhanced density caused by compression. 
\item We find that the gas mass in this warm phase is between 40-76\% of the mass stripped from the satellite and deposited in the wakes
($M_{\rm warm}\sim1.4-7.6\times10^9$M$_\odot$ for the parameters we have used in our simulations). This gas is mostly coming from the 
satellite, but a non-negligible fraction, 20-28\%, comes from the ICM.
\item The analytic model described in Section \ref{theo} is able to roughly estimate the amount of removed mass
and the time scale of removal due to the two main
acting mechanisms: ram pressure stripping and KH instabilities. Moreover,
once properly calibrated, it can be used to estimate the most likely temperature and cooling time of the gas mixture after 
the gas settles in the turbulent wakes. We believe that this model can be incorporated into semi-analytic models of galaxy formation to 
account for these processes, which are typically ignored in current implementations.
\end{itemize}

We speculate that the warm phase created through the mechanism proposed in this work has the potential to become a mode of star 
formation in galaxy groups (through in situ star formation and/or through a cold stream flowing to the central galaxy of the host). 
In fact, the possibility of stars forming in turbulent wakes from the condensation of gas stripped from cold discs seems to be 
a likely hypothesis in some observed cases \citep[e.g.][]{Hester_2010}. Although in such cases the mechanism at work is the same we
have studied here, star formation is far more likely in those since the gas being removed is much colder
than the ambient gas. We are thus cautious in speculating about the fate of the removed coronal gas. 
In particular, it is not clear whether the gas mixture can survive long enough without evaporating through conduction with the ICM 
\citep[e.g.][]{Nipoti_2004}. This would depend on the specific details of conduction in the surfaces of the turbulent wake, which are difficult
to estimate. 
It is therefore crucial to address these questions using more detailed simulations. We believe that the results of this work motivate  
further analyses to fully quantify the effect of the mixing of the coronal gas from satellites with the intragroup gas.

\section*{Acknowledgments}

We thank the anonymous referee for helpful suggestions.
We thank John ZuHone for helpful comments on the simulation setup with FLASH. We 
acknowledge David Gilbank and 
Ting Lu for helpful conversations during the early stages of this work, and Volker Springel and Mark Vogelsberger for
helpful discussions towards the completion of this paper. NA wishes to thank David Spergel for helpful discussions.
JZ and NA are supported by the University of 
Waterloo and the Perimeter Institute for Theoretical Physics. Research at Perimeter Institute is supported by the Government of Canada 
through Industry Canada and by the Province of Ontario through the Ministry of Research \& Innovation. JZ acknowledges financial support by a 
CITA National Fellowship. MLB acknowledges a NSERC Discovery Grant. This work was made possible by the facilities of the Shared Hierarchical 
Academic Research Computing Network (SHARCNET:www.sharcnet.ca) and Compute/Calcul Canada.

\appendix

\section{RPS time scale}\label{app_1}

The properties of the initial shock generated by the colliding coronae and the values of the post-shock thermodynamic variables 
can be estimated by solving a one-dimensional shock tube problem. Once the shock develops, let region 1 to the left be the 
unperturbed ICM moving to the right with a relative velocity $v_{\rm sat}$ and with properties 
$(\rho, P, u, c_{\rm s})_1=(\rho_{\rm host}, P_{\rm host}, v_{\rm sat}, c_{\rm s, host})$, and region 4 to the right be the unperturbed 
corona of the satellite at rest having properties: 
$(\rho, P, u, c_{\rm s})_4=(\rho_{\rm sat}(r_{\rm v}), P_{\rm sat}(r_{\rm v}), 0, c_{\rm s, sat}(r_{\rm v}))$. Regions 1 to 4 separate the three
{\it characteristics} of the problem. The forward shock front divides regions 3 and 4; thus, region 3 contains the post-shock
material. The solution to this problem is found by solving the following transcendental equation:
\begin{equation}\label{trans}
  \frac{P_{1}}{P_4}=1=\xi\left[1+\frac{\gamma-1}{c_{\rm s, 1}}\left(u_1-\frac{c_{\rm s,4}}{2\gamma}
    \frac{\xi-1}{\sqrt{\lambda(\xi-1)+1}}\right)\right]^{-2\gamma/(\gamma-1)},
\end{equation}
where $\xi=P_3/P_4$ and $\lambda=(\gamma+1)/2\gamma$, and we have assumed that the corona of the satellite and the ICM are in 
pressure equilibrium at first: $P_{1}/P_{4}=1$. The velocity of the shock, and the velocity of the post-shock fluid are then given by:
\begin{eqnarray}\label{speed}
  v_{\rm sh}&=&c_{\rm s,4}\sqrt{\lambda(\xi-1)+1},\nonumber\\
  v_{\rm ps}&=&c_{\rm s,4}\frac{(\xi-1)}{\gamma\sqrt{\lambda(\xi-1)+1}}.
\end{eqnarray}
Once the shock is generated it will propagate towards the centre of the satellite. The density stratification of the latter will 
then change the speed of the shock and modify the post-shock properties of the newly encountered material. To roughly estimate
the propagation of the shock, we follow the description of \citet{Kogure_1962} that reduces the evolution of the 
post- and pre-shock pressure ratio $\xi$ to the following
ordinary differential equation (derived from a one-dimensional problem neglecting the effects of gravity and radiative losses, 
see their Eq. 3.1):
\begin{eqnarray}\label{diff}
  \frac{d\xi}{ds'}&=&-\frac{(1+\xi)(1+\lambda\xi)(2+(3-\lambda)\xi)}{4+2(2+\lambda)\xi+\lambda(3-\lambda)\xi^2}\frac{d{\rm ln}P}{ds'}\nonumber\\
  &&+\frac{\xi(1+\xi)(1+\lambda\xi)}{4+2(2+\lambda)\xi+\lambda(3-\lambda)\xi^2}\frac{d{\rm ln}\rho}{ds'},
\end{eqnarray}
where $s'$ is the Lagrange coordinate of the shock front. In reference to the variable $s=r/r_{\rm v}$ from the centre of the satellite
in Eqs.~(\ref{nfw}) and (\ref{hse}), $s'=1-s$. Thus, solving Eq.~(\ref{trans}) assuming pressure equilibrium in the boundary
we get $\xi(s'=0)=9.0 (12.9)$ for the galaxy-group (galaxy-cluster) interaction. With this boundary condition, 
we can solve Eq.~(\ref{diff}) to find $\xi(s')$ and use it in the first of Eqs.~(\ref{speed}) to get the shock speed as a function 
of $s'$, which can then be substituted in Eq.~(\ref{t_rps}) to find the time scale for RPS.

\bibliography{lit}

\end{document}